\documentclass[journal,twocolumn,10pt]{IEEEtran}
\usepackage{graphicx}
\usepackage{amssymb}
\usepackage{mathtools}
\usepackage{dsfont}
\usepackage{cite}
\usepackage{stfloats}
\usepackage{subfigure}
\usepackage{psfrag}
\usepackage[mathscr]{euscript}
\usepackage{acronym}  
\usepackage{amsmath}
\usepackage{algorithm}
\usepackage[noend]{algpseudocode}
\usepackage{booktabs}
\usepackage{bbm}

\makeatletter
\def\BState{\State\hskip-\ALG@thistlm}
\makeatother

\usepackage{float}
\usepackage{balance}

\acrodef{CCDF}{complementary cumulative distribution function}
\acrodef{CF}{characteristic function}
\acrodef{PPP}{Poisson point processe}
\acrodef{RV}{random variable}
\acrodef{i.i.d.}{independent and identically distributed}
\acrodef{PDF}{probability distribution function}
\acrodef{CDF}{cumulative distribution function}
\acrodef{ch.f.}{characteristic function}
\acrodef{AWGN}{additive white Gaussian noise}
\acrodef{SNR}{signal-to-noise ratio}
\acrodef{LRT}{likelihood ratio test}
\acrodef{DRT}{distance ratio test}
\acrodef{GLRT}{generalized likelihood ratio test}
\acrodef{CRLB}{Cram\'{e}r-Rao lower bound}
\acrodef{CRB}{Cram\'{e}r-Rao bound}
\acrodef{ZZLB}{Ziv-Zakai lower bound}
\acrodef{ZZB}{Ziv-Zakai bound}
\acrodef{LOS}{line-of-sight}
\acrodef{ToF}{time-of-flight}
\acrodef{NLOS}{non-line-of-sight}
\acrodef{GDOP}{geometric dilution of precision}
\acrodef{GPS}{Global Positioning System}
\acrodef{FIM}{Fisher information matrix}
\acrodef{PEB}{position error bound}
\acrodef{SPEB}{squared position error bound}
\acrodef{TOA}{time-of-arrival}
\acrodef{TOF}{time-of-flight}
\acrodef{WSN}{wireless sensor network}
\acrodef{MAC}{medium access control}
\acrodef{RSS}{received signal strength}
\acrodef{WAF}{wall attenuation factor}
\acrodef{TDOA}{time difference-of-arrival}
\acrodef{RF}{radiofrequency}
\acrodef{RTT}{round-trip time}
\acrodef{AOA}{angle-of-arrival}
\acrodef{MF}{matched filter}
\acrodef{ED}{energy detector}
\acrodef{ML}{maximum likelihood}
\acrodef{MSE}{mean-square error}
\acrodef{RMSE}{root-mean-square error}
\acrodef{LEO}{localization error outage}
\acrodef{ppm}{part-per-million}
\acrodef{ACK}{acknowledge}
\acrodef{UWB}{Ultrawide bandwidth}
\acrodef{TNR}{threshold-to-noise ratio}
\acrodef{LS}{least squares}
\acrodef{IR-UWB}{impulse radio UWB}
\acrodef{FCC}{Federal Communications Commission}
\acrodef{TH}{time-hopping}
\acrodef{PPM}{pulse position modulation}
\acrodef{MUI}{multi-user interference}
\acrodef{PDP}{power delay profile}
\acrodef{BPZF}{band-pass zonal filter}
\acrodef{SIR}{signal-to-interference ratio}
\acrodef{SINR}{signal-to-interference-plus-noise ratio}
\acrodef{RFID}{radio frequency identification}
\acrodef{WPAN}{wireless personal area network}
\acrodef{WWB}{Weiss-Weinstein bound}
\acrodef{DP}{direct path}
\acrodef{MF}{matched filter}
\acrodef{MMSE}{minimum-mean-square-error}
\acrodef{SBS}{serial backward search}
\acrodef{SBSMC}{serial backward search for multiple clusters}
\acrodef{NBI}{narrowband interference}
\acrodef{WBI}{wideband interference}
\acrodef{INR}{interference-to-noise ratio}
\acrodef{CR}{channel response}
\acrodef{CIR}{channel impulse response}
\acrodef{CR}{channel  response}
\acrodef{RADAR}{radar}
\acrodef{MUR}{Multistatic radar}
\acrodef{JBSF}{jump back and search forward}
\acrodef{HDSA}{high-definition situation-aware}
\acrodef{RRC}{root raised cosine}
\acrodef{ST}{simple thresholding}
\acrodef{BTB}{Bellini-Tartara bound}
\acrodef{P-Max}{$P$-Max}  
\acrodef{MIMO}{multiple-input multiple-output}
\acrodef{MAP}{maximum a posteriori}
\acrodef{FG}{factor graph}
\acrodef{OP}{outage probability}
\acrodef{WED}{wall extra delay}
\acrodef{RMS}{root mean square}
\acrodef{SPAWN}{sum-product algorithm over a wireless network}
\acrodef{MDD}{minimum distance distribution}
\acrodef{MAP}{maximum a posteriori probability}
\acrodef{SAP}{small cell access point}
\acrodef{UE}{user equipment}
\acrodef{MBS}{macro cell base station}
\acrodef{UER}{\ac{UE} Relay}
\acrodef{D2D}{device-to-device}
\acrodef{MBS}{macro base station}
\acrodef{CSI}{channel state information}
\acrodef{OGR}{outage guard region}
\acrodef{FUR}{feasible UER region}
\acrodef{EHR}{energy harvesting region}
\acrodef{EH}{energy harvesting}
\acrodef{D2D-EHSN}{D2D communication provided \ac{EH} small cell network}
\acrodef{D2D-EHHN}{D2D communication provided \ac{EH} heterogeneous network}
\acrodef{3GPP}{3rd Generation Partnership Project}
\acrodef{BS}{base station}
\acrodef{DF}{decode and forward}
\acrodef{CCDF}{complementary cumulative distribution function}
\acrodef{ZF}{zero forcing}
\acrodef{RZF}{regularized zero forcing}
\acrodef{WLLN}{weak law of large number}
\acrodef{SLLN}{strong law of large numbers}
\acrodef{TDD}{Time-division duplex}
\acrodef{EE}{energy efficiency} 
\acrodef{HetNet}{heterogeneous network} 
\acrodef{SCP}{Single Cell Processing}
\acrodef{CBF}{Coordinated Beamforming}
\usepackage{color}
\usepackage{dsfont}
\usepackage{bbm}

\DeclareMathAlphabet{\mathsf}{OML}{cmbr}{m}{it}

\newtheorem{definition}{\bf Definition}
\newtheorem{theorem}{\bf Theorem}
\newtheorem{lemma}{\bf Lemma}
\newtheorem{corollary}{\bf Corollary}

\newtheorem{remark}{\bf Remark}

\newtheorem{assumption}{\bf Assumption}

\newcommand{\bd}{\begin{description}}
\newcommand{\ed}{\end{description}}
\newcommand{\be}{\begin{enumerate}}
\newcommand{\ee}{\end{enumerate}}
\newcommand{\bi}{\begin{itemize}}
\newcommand{\ei}{\end{itemize}}
\newcommand{\bl}{\begin{list}}
\newcommand{\el}{\end{list}}
\newcommand{\bt}{\begin{tabbing}}
\newcommand{\et}{\end{tabbing}}

\newcommand{\paperTitle}{ Spatiotemporal Analysis for Age of Information in Random Access Networks under Last-Come First-Serve with Replacement Protocol }

\begin{document}

{
\title{\paperTitle}

\author{

	    Howard~H.~Yang, \textit{Member, IEEE},
        Ahmed Arafa, \textit{Member, IEEE},\\
	    Tony~Q.~S.~Quek, \textit{Fellow, IEEE},
	    and H.~Vincent~Poor, \textit{Fellow, IEEE}


\thanks{H. H. Yang is with the Zhejiang University/University of Illinois at Urbana-Champaign Institute, Zhejiang University, Haining 314400, China, the College of Information Science and Electronic Engineering, Zhejiang University, Hangzhou 310007, China, and the Department of Electrical and Computer Engineering, University of Illinois at Urbana-Champaign, Champaign, IL 61820, USA (email: haoyang@intl.zju.edu.cn).

A.~Arafa is with the Department of Electrical and Computer Engineering, University of North Carolina at Charlotte, NC 28223, USA (email: aarafa@uncc.edu).

T.~Q.~S.~Quek is with the Information Systems Technology and Design Pillar, Singapore University of Technology and Design, Singapore 487372 (e-mail: tonyquek@sutd.edu.sg).

H.~V.~Poor is with the Department of Electrical Engineering, Princeton University, Princeton, NJ 08544 USA (e-mail: poor@princeton.edu).
}
}
}
\maketitle
\acresetall
\thispagestyle{empty}
\begin{abstract}
We investigate the age-of-information (AoI) in the context of random access networks, in which transmitters need to send a sequence of information packets to the intended receivers over a shared spectrum. Due to interference, the dynamics at the link pairs will interact with each other over both space and time, and the effects of these spatiotemporal interactions on the AoI are not well understood.
In this paper, we straddle queueing theory and stochastic geometry to establish an analytical framework, that accounts for the interplay between the temporal traffic attributes and spatial network topology, for such a study.
Specifically, we derive accurate and tractable expressions to quantify the network average AoI as well as the outage probability of peak AoI.
Besides, we develop a decentralized channel access policy that exploits the local observation at each node to make transmission decisions that minimize the AoI.
Our analysis reveals that when the packet transmissions are scheduled in a last-come first-serve (LCFS) order, whereas the newly incoming packets can replace the undelivered ones, depending on the deployment density, there may or may not exist a tradeoff on the packet arrival rate that minimizes the network average AoI. Moreover, the slotted ALOHA protocol is shown to be instrumental in reducing the AoI when the packet arrival rates are high, yet it cannot contribute to decreasing the AoI in the regime of infrequent packet arrivals.
The numerical results also confirm the efficacy of the proposed scheme, where the gain is particularly pronounced when the network grows in size because our method is able to adapt the channel access probabilities with the change of ambient environment.
\end{abstract}
\begin{IEEEkeywords}
Poisson bipolar network, age of information, channel access probability, queueing theory, stochastic geometry.
\end{IEEEkeywords}

\acresetall

\section{Introduction}\label{sec:intro}
The age-of-information (AoI) is a metric that measures the ``freshness'' of information packets delivered over a period of time \cite{KauYatGru:12}, which has been used for the design of networking schemes to provide timely status updates for real-time applications \cite{KosPapAng:17}.
Compared with the transmitter-centric metrics, e.g., delay or throughput, AoI is usually adopted at the receiver side to measure the time elapsed since the generation of the latest delivered packet, thus being able to gauge the “freshness” associated with the information packets \cite{KosPapAng:17,SunUysYat:17,YatKau:18,ZhaAraHua:19,AraYanUlu:18,WuYanWu:18,BacSunUys:19}.
As such, networks designed by minimizing the metric of AoI enable the acquisition of fresh data and are particularly relevant in the Internet of Things (IoT) applications where the timeliness of information is crucial, e.g., monitoring the status of a system or asserting remote controls based on information collected from a network of sensors \cite{AbdPapDhi:19,ZhoSaa:18,XuYanWan:20,XuWanYan:20INFOCOM}.

Because these platforms generally constitute a random access network in which multiple source nodes need to communicate with their destinations via a shared spectrum, the interference amongst transmitters located in geographical proximity may be severe and lead to transmission failures that hinder the timely updates of information.
In response, a number of strategies to schedule the set of simultaneously active links have been proposed \cite{YatKau:17ISIT,YavUys:21,Mun:21,ZhoSaa:20a,ZhoSaa:20b,HeYuaEph:16,talak2018optimizing,BedSunKom:19,HsuModDua:19,KadSinUys:18,KadSinMod:19}, aiming to minimize the information age by pertaining the interference to an acceptable range.
Particularly, when a large number of devices are updatign their time-stamped status to a common receiver over a multiple access channel, \cite{YatKau:17ISIT} analyzed the AoI under two classical channel access policies, i.e., the round robin and slotted ALOHA, and showed that round robin outperforms slotted ALOHA in terms of AoI.  
Nevertheless, it is shown that the slotted ALOHA can be modified to minimize AoI by either introducing an age threshold \cite{YavUys:21} or operating it in an irregular repetition manner \cite{Mun:21}. 
If the devices are scheduled under a carrier sense multiple access (CSMA)-type channel access, \cite{ZhoSaa:20a} provides closed-form expressions for obtaining insights to the average AoI and average peak AoI, as well as a mean-field game framework that optimizes the channel access to minimize the AoI, which is robust even under noisy channels \cite{ZhoSaa:20b}. 
On the other hand, in random access networks where each source is paired with a dedicated receiver and transmissions over the same channel collide, \cite{HeYuaEph:16} demonstrated that the link scheduling problem for AoI minimization is NP-hard and proposed a steepest age descent algorithm to solve the problem in a suboptimal but fast manner.
It is further shown that the optimal scheduling policy and the optimal sampling strategy can be independently devised \cite{talak2018optimizing,BedSunKom:19}, followed by a variety of scheduling schemes, ranging from randomized \cite{KadSinMod:19} or index based \cite{BedSunKom:19,KadSinUys:18,KadSinMod:19}, to using the structural Markov decision process \cite{HsuModDua:19}.
Furthermore, several threshold-based channel access schemes have also been developed to optimize the AoI from a network perspective \cite{CheGatHas:19,CheGuLie:20}.
However, these results are devised based on collision models or conflict graphs, which do not precisely capture the key attributes of a wireless system.
Indeed, transmissions over the spectrum are entangled in a slew of physical factors, particularly the fading, path loss, and co-channel interference.
As such, it is suggested to adopt the signal-to-interference-plus-noise ratio (SINR) model rather than the conflict graph for a better characterization of the source-destination communication processes so as to obtain genuine understanding of the insights \cite{HeYuaEph:16,SanBac:17}.
Recognizing this, a recent line of research has been carried out \cite{HuZhoZha:18,ManAbdDhi:20,EmaElSBau:20,YanArafaQue:19,YanXuWan:19,ManCheAbd:20b}, that conflates queueing theory with stochastic geometry -- a disruptive tool for assessing the performance of wireless links in large-scale networks -- to account for the spatial, temporal, and physical level attributes in the analysis of AoI. 
Such spatiotemporal analyses are generally challenging because the evolution of queues associated with the transmitters are coupled with each other over space and time via the interference they caused. 
In response, \cite{HuZhoZha:18} adopts the favorable/dominant system argument to decouple the spatial-temporal correlations and derive lower and upper bounds for the distribution of average AoI in the context of a Poisson network. 
Additionally, via a careful (re)construction of the dominant system, tighter upper bounds for the spatial distribution of mean peak AoI is derived for a large system under both preemptive and non-preemptive queueing disciplines \cite{ManAbdDhi:20}.
Furthermore, based on the dominant system, a distributed algorithm that configures the channel access probabilities at each individual transmitter based on the local observation of the network topology is proposed to minimize the peak AoI \cite{YanArafaQue:19}.
To obtain more accurate expressions for AoI rather than the bounds, \cite{EmaElSBau:20} resort to the meta distribution and evaluate the performance of peak AoI in uplink IoT networks under time-triggered and event-triggered traffic profiles. 
And \cite{YanXuWan:19} improves the analysis of AoI by characterizing the spatial dependency amongst the transmitters by modeling the locations of interfering nodes as an inhomogeneous Poissoin point process (PPP). 
Moreover, \cite{ManCheAbd:20b} considers a generate at will model of the transmitters and provides a joint spatio-temporal analysis of AoI and throughput for cellular-based IoT networks with heterogeneous traffic. 
Nonetheless, \cite{EmaElSBau:20,YanArafaQue:19,YanXuWan:19} schedule the transmissions of information packets in a first-come first-serve (FCFS) discipline where the failed packets are perpetually retransmitted till successfully received, which is not appealed for minimizing the AoI. Although \cite{YanXuWan:19} also investigated the last come first serve with preemption (LCFS-PR) discipline, the transmitters still maintain an infinite-size queue to store the incoming packets and resend the undelivered packets during the available time slots, which introduced unnecessary interference. 
As pointed out by \cite{CheGatHas:19}, under the metric of AoI, it is implicitly assumed that the information content of the packets form a Markov process. Therefore, an AoI-optimal transmission protocol shall have the transmitters discard the older undelivered packets upon the arrival of fresher packets so as to reduce interference and enhance the probability of successful transmissions. 
Under such a protocol, the only relevant spatiotemporal analysis of AoI is provided in \cite{ManAbdDhi:20}, but the performance metric considered in that work is the \textit{peak} AoI while the more commonly used metric of \textit{average} AoI has not been explored. 
Besides, whether a locally adaptive scheme can be devised to further minimize the AoI is also not clear. 


\subsection{Approach and Summary of Results}
In this paper, we aim at developing a theoretical template for a thorough understanding of the AoI over a random access network. On that purpose, we model the positions of transmitter-receiver pairs as a Poisson bipolar network. Each transmitter generates a sequence of status updates, encapsulated in the information packets, according to independent Bernoulli processes.
The newly incoming packets at each transmitter are stored in a unit-size buffer and replace the older undelivered ones, if any.
In each time slot, transmitters with non-empty buffers employ a slotted ALOHA protocol to access the shared spectrum and send out packets.
The transmissions are successful only if the received SINR exceeds a decoding threshold, upon which the packet can be removed from the transmitter buffer.
Otherwise, the packet stays in the buffer and will be retransmitted in the next available time slot (unless replaced by a newly generated packet).
Because of interference, there are coupling effects amongst the node positions and their buffer states. 
For tractable analyses, we adopt the mean-field approximation to decouple such spatial-temporal correlations, and jointly use tools from stochastic geometry, to capture the macroscopic interference behavior, and queueing theory, to characterize the evolution of queues at the microscopic level, to derive accurate expressions for both the peak and average AoI.
Leaning on the theoretical framework, we leverage similar techniques as \cite{YanArafaQue:19} to devise a locally adaptive slotted ALOHA protocol to minimize the AoI.
The analytical results enable us to explore the effects of different network parameters on the AoI performance and hence can serve as useful guidelines for further system designs.
Our main contributions are summarized below.
\begin{itemize}
\item We establish a mathematical framework for the analysis of AoI in random access networks. Our model is general and encompasses several key features of a wireless system, including the channel gain, path loss, deployment density, interference, and spatially queueing interactions.
\item We derive accurate expressions for the average AoI as well as the outage probability of peak AoI. By resorting to different special cases, we can obtain simple expressions  from the analysis that facilitate an intuitive understanding of the AoI in random access networks.
\item Building upon the theoretical framework, we develop a locally adaptive slotted ALOHA protocol, which exploits the local observation at each node to configure a link-wise channel access probability that minimizes the AoI across the network. The proposed scheme is fully decentralized and has a low implementation complexity.
\item Numerical results reveal that: $i$) when the wireless links are densely deployed, there exists an optimal update frequency that minimizes the network average AoI or the outage probability of peak AoI, while in a sparsely deployed network, the AoI monotonically decreases with the packet arrival rate, $ii$) the slotted ALOHA protocol is only effective when the packet arrival rates are high (and/or the topology is dense) and it cannot contribute to reducing the AoI in regimes of infrequent packet arrivals, and $iii$) the proposed channel access policy is able to maintain the network average AoI at a low level for a wide range of deployment density because it can adjust the frequency of radio channel access with the change of ambient environment.
\end{itemize}

The remainder of the paper is organized as follows. We introduce the system model in Section II. In Section III, we detail the analysis of the average and peak AoI, and provide a series of discussions for insights. We show the simulation and numerical results in Section IV, which confirm the accuracy of our analysis and provide insights about the AoI performance of a large-scale wireless network. We conclude the paper in Section V.

\begin{table}
\caption{Notation Summary
} \label{table:notation}
\begin{center}
\renewcommand{\arraystretch}{1.3}
\begin{tabular}{c  p{5.5cm} }
\hline
{\bf Notation} & {\hspace{2.5cm}}{\bf Definition}
\\
\hline
$\tilde{\Phi}$; $\lambda$ & PPP modeling the locations of transmitters; transmitter spatial deployment density \\
$\bar{\Phi}$; $\lambda$ & PPP modeling the locations of receivers; receiver spatial deployment density \\
${\Phi}$ & Superposition of PPPs $\tilde{\Phi}$ and $\bar{\Phi}$, i.e., $\Phi = \tilde{\Phi} \cup \bar{\Phi}$ \\
$P_{\mathrm{tx}}$; $\sigma^2$ &  Transmit power; power of the thermal noise \\
$\rho$; $\alpha$ &  Signal-to-noise ratio; path loss exponent \\
$\xi$; $\theta$ & Packet update frequency; SINR decoding threshold \\
$r$; $p$ & Distance of a transmitter-receiver pair; slotted ALOHA channel access probability \\
$\mu_{0, t}^\Phi$ & Transmission success probability of link $0$ at time slot $t$, conditioned on the point process $\Phi$ \\
$a_j^\Phi$ & Buffer non-empty probability at transmitter node $j$, conditioned on the point process $\Phi$ \\
$W$  & Local observation window of the transmitters, given in the form of a stopping set \\
$\gamma^\Phi_j$ & Probability of channel access at transmitter $j$, constructed based on the local information \\
$\bar{\Delta}$; $\hat{\Delta}$ & Network average AoI; network peak AoI \\
\hline
\end{tabular}
\end{center}\vspace{-0.63cm}
\end{table}%

\begin{figure*}[t!]
  \centering{}

    {\includegraphics[width=1.95\columnwidth]{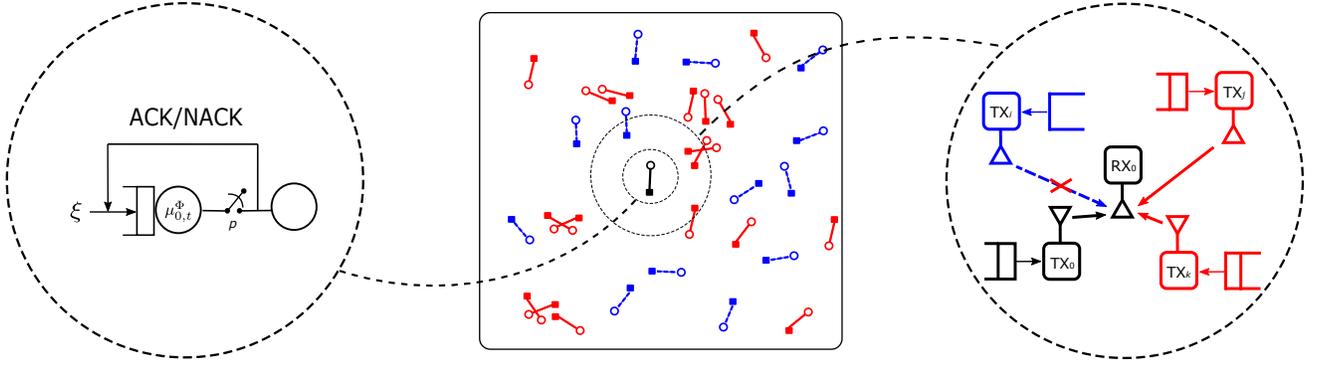}}

  \caption{The employed wireless network at microscopic and macroscopic scales, where the squares and circles denote the transmitters and receivers, respectively: the typical link is the black solid line, the other active links are denoted by the red solid lines, and the inactive links are the dashed lines in blue. }
  \label{fig:NetwMod}
\end{figure*}

\section{System Model}\label{sec:sysmod}
In this section, we introduce the setup of the network model, as well as the concepts of average and peak AoI. 
The main notations used throughout the paper are summarized in Table~\ref{table:notation}.
\subsection{Spatial Configuration and Physical Layer Parameters }
Let us consider a wireless network, as depicted in Fig.~\ref{fig:NetwMod}, that consists of a set of transmitter-receiver pairs, all located in the Euclidean plane.
The transmitters are scattered according to a homogeneous Poisson point process (PPP) $\tilde{\Phi}$ of spatial density $\lambda$, where a generic node $i$ located at $X_i \in \tilde{\Phi}$ has one dedicated receiver at $y_i$, which is at distance $r$ from $X_i$ and oriented in a uniformly random direction.\footnote{Such a setting is commonly known as the Poisson bipolar model \cite{BacBla:09}, which is a large-scale analog to the classical model of \textit{Random Networks} \cite{GupKum:00} and has been widely used for the modeling of networks without a centralized controller, e.g., the D2D, IoT, and wireless ad-hoc networks.
Note that the analysis developed in this paper can be extended to investigate the AoI in cellular networks.}
According to the displacement theorem \cite{BacBla:09}, the location set $\bar{\Phi} = \{y_i\}_{i=0}^\infty$ also forms a homogeneous PPP with spatial density $\lambda$.
If a transmitter needs to communicate with its receiver, it employs a constant power\footnote{We unify the transmit power for tractability, while the framework developed in this paper can be used to study the effects of power control in similar spirits to \cite{RamBac:19,ChiElSCon:19}. } $P_{\mathrm{tx}}$ and sends out packets over a shared spectrum, which is affected by small-scale fading that follows Rayleigh distribution with a unitary mean and large-scale path-loss that follows power law attenuation.
All channel gains are independent and identically distributed (i.i.d.) across space and time.
Besides, the transmission is also subject to Gaussian thermal noise with a total variance $\sigma^2$.

\subsection{Temporal Configuration and Transmission Protocol }
We assume the network is synchronized\footnote{Synchronization over networks can be achieved by either centralized \cite{WanLinAdh:17} or distributed mechanisms \cite{XioWuShe:17}. } and the time is segmented into equal-duration intervals, which are referred to as time slots. We further assume the transmission of each packet occupies exactly one time slot.
At the beginning of each time slot, every transmitter has an arrival of information packet with probability $\xi \in (0,1]$.
The newly incoming packet at each transmitter will be stored in a unit-size buffer and replace the undelivered older one if there is any.
In that respect, the schedule of packet transmissions constitutes a last-come first-serve with replacement (LCFS-R) protocol.

In each time slot, transmitters with non-empty buffers adopt the slotted ALOHA protocol with probability $p \in (0, 1)$ to access the radio channel and send out packets.
A transmission is considered successful if the SINR received at the destination exceeds a decoding threshold, upon which the receiver sends an ACK feedback message so that the packet can be removed from the buffer.
Otherwise, the receiver sends a NACK feedback message and the packet is retransmitted in the next available time slot.\footnote{We assume the ACK/NACK transmission is instantaneous and error-free, as commonly done in the literature \cite{TalKarMod:18}.}
In this network, the delivery of packets incurs a delay of one time slot, namely, packets are transmitted at the beginning of time slots and, if the transmission is successful, they are delivered by the end of the same time slot.

Because the time scale of fading and packet transmission is much smaller than that of the spatial dynamics, we assume the network topology is static, i.e., an arbitrary but fixed point pattern is realized at the beginning and remains unchanged over the time domain.

\subsection{Age of Information}
The performance metric of this work is the AoI, which captures the timeliness of information delivered at the receiver side.
A formal definition of this metric is stated in below.
\begin{definition}
\textit{Consider a typical transmitter-receiver pair. Let $\{G(t_i)\}_{i \geq 1}$ be the sequence of generation times of information packets that were delivered and $\{t_i\}_{ i \geq 1 }$ be the corresponding times at which these packets are received at the destination. Amongst the packets received till time $t$, denote the index of the latest generated one by $n_t = \arg\max_{i}\{ G(t_i) | t_i \leq t \}$. The age of information at the receiver is defined as $\Delta(t) = t - G(t_{n_t})$.
}
\end{definition}

If the average time for packet delivery is the same, then according to Definition~1, the presence of a new packet at the transmitter will make the older one irrelevant in reducing the AoI.
As such, maintaining a unit-size buffer at each transmitter and replacing undelivered packets with newly incoming ones is consistent with the minimum AoI packet management strategy.

\begin{figure}[t!]
  \centering{}

    {\includegraphics[width=0.95\columnwidth]{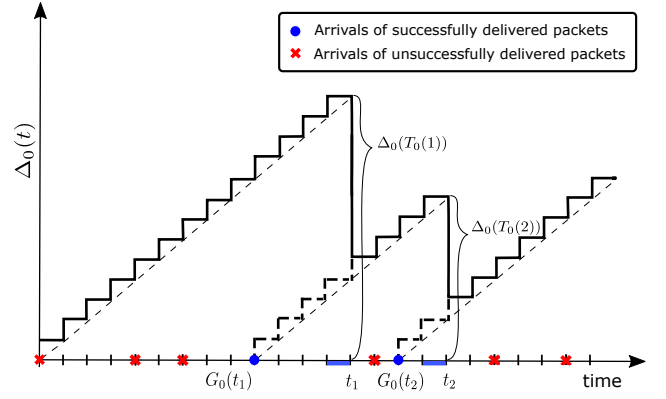}}

  \caption{ AoI evolution example at a typical link under the LCFS-R discipline. The time instances $G_0(t_i)$ and $t_i$ respectively denote the moments when the $i$-th packet is generated and delivered, and the age is reset to $t_i - G_0(t_i)$. Here, $t_i = T_0(i)$ with $T_0(i)$ defined in \eqref{equ:DefPAoI}. }
  \label{fig:AoIMod_V1}
\end{figure}

Without loss of generality, we denote the link pair located at $(X_0, y_0)$, where $y_0$ is the origin, as typical.
Note that although the considered wireless network contains infinitely many dipoles, thanks to the stationary property of PPP, the AoI of each wireless link is statistically equivalent.
Under the employed system model, the AoI of the typical link goes up by one in each time slot if no new packet is updated at the receiver side, and, when the update is received, reduces to the time elapsed since the generation of the delivered packet. An example of the dynamics of AoI is illustrated in Figure~\ref{fig:AoIMod_V1}.
Formally, the evolution of $\Delta_0(t)$ can be written as follows:
\begin{align*}
\Delta_0(t \!+\! 1) =
\left\{
       \begin{array}{ll}
         \!\!  \Delta_0(t)  +   1, \quad \quad ~~   \text{if no update received}, \\
         \!\!  t  -  G_0(t) + 1, \quad \quad ~~ \text{otherwise}
       \end{array}
\right.
\end{align*}
where $G_0(t)$ is the generation time of the packet delivered over the typical link by the end of time slot $t$.

In this work, we leverage two quantities, namely the \textit{average} and \textit{peak} AoI, as our metric to evaluate the freshness of information over a random access network. Specifically, the average AoI at a given link $j$ is defined as
\begin{align}
\bar{\Delta}_j = \limsup_{ T \rightarrow \infty } \frac{ 1 }{ T } \sum_{ t=1 }^T \Delta_j(t),
\end{align}
and the peak AoI is
\begin{align} \label{equ:DefPAoI}
\hat{\Delta}_j = \limsup\limits_{ N \rightarrow \infty } \frac{ \sum_{n=1}^N \Delta_j( T_j(n) ) }{N},
\end{align}
where $T_j(n)$ is the time slot at which the $n$-th packet from transmitter $j$ is successfully delivered to the receiver. By extending these concepts to the context of a network, we define the \textit{network} average and peak AoI respectively as follows:
\begin{align}
\bar{ \Delta } &= \limsup_{R \rightarrow \infty} \frac{ \sum_{ X_j \in \tilde{\Phi} \cap B(0,R) } \bar{\Delta}_j }{ \sum_{ X_j \in \tilde{\Phi} } \mathbbm{1}\{ X_j \!\in\! B(0,R) \}  }
\nonumber\\
&\stackrel{(a)}{=}   \mathbb{E}^0 \Big[ \limsup_{ T \rightarrow \infty } \frac{1}{T} \sum_{t=1}^T \Delta_0( t )  \Big]
\end{align}
and
\begin{align}
\hat{\Delta} &= \limsup_{R \rightarrow \infty} \frac{ \sum_{ X_j \in \tilde{\Phi} \cap B(0,R) } \hat{\Delta}_j }{ \sum_{ X_j \in \tilde{\Phi} } \mathbbm{1}{ \{ X_j \!\in\! B(0,R) \} }  }
\nonumber\\
& = \mathbb{E}^0 \Big[ \limsup_{ N \rightarrow \infty } \frac{1}{N} \sum_{n=1}^N \Delta_0( T_0(n) )  \Big],
\end{align}
where $B(0,R)$ denotes a disk centered at the origin with radius $R$, $\mathbbm{1\{ \cdot \}}$ is the indicator function, and $(a)$ follows from the Campbell's theorem \cite{BacBla:09}. The notation $\mathbb{E}^0[\cdot]$ indicates the expectation is taken with respect to the Palm distribution $\mathbb{P}^0$ of the stationary point process where under $\mathbb{P}^0$ almost surely there is a node located at the origin \cite{BacBla:09}.

\section{ Analysis }
This section constitutes the main technical part of our paper, in which we derive analytical expressions to characterize the statistics of AoI.
Specifically, we analyze the distribution of packet depletion rate, or equivalently the conditional transmission success probability, at each communication link.
Based on that, we calculate the value of average AoI, as well as the outage probability of peak AoI, of the considered wireless network. For better readability, most proofs and mathematical derivations have been relegated to the Appendix.
\subsection{Preliminaries}
\subsubsection{SINR at a typical receiver}
Due to the stationary property of PPPs, we can apply Slivnyak's theorem \cite{BacBla:09} and concentrate on a \textit{typical receiver} located at the origin, with its tagged transmitter situated at $X_0$. Note that when averaging over the point process, this representative link has the same statistic as those obtained by averaging over other links in the network.
As such, if the transmitter sends out a packet during time slot $t$, the SINR received at the destination can be written as
\begin{align} \label{equ:SINR_expression}
\mathrm{SINR}_{0,t} = \frac{P_{\mathrm{tx}} H_{00} r^{-\alpha} }{ \sum_{ j \neq 0 } P_{\mathrm{tx}} H_{j0} \zeta_{j,t} \nu_{j,t} \Vert X_j \Vert^{-\alpha} + \sigma^2 }
\end{align}
where $\alpha$ denotes the path loss exponent, $H_{ji} \sim \exp(1)$ is the channel fading from transmitter $j$ to receiver $i$, $\zeta_{j,t} \in \{ 0, 1 \}$ is an indicator showing whether the buffer of node $j$ is empty ($\zeta_{j,t}=0$) or not ($\zeta_{j,t}=1$), and $\nu_{j,t} \in \{ 0, 1 \}$ represents the state of channel access at node $j$, where it is set to 1 upon assuming transmission approval and 0 otherwise.
\subsubsection{Conditional transmission success probability}
Since the information packets are generated according to independent Bernoulli processes, seen from the temporal perspective, the interval of packet arrivals at any given link follows a geometric distribution. However, due to interference, the packet transmission process of the same wireless link has a rate -- often characterized by the transmission success probability -- that is dependent on the network topology as well as the buffer states of the other nodes. To that end, the distribution of packet departures intervals is generally unknown. By noticing that each transmitter maintains a unit-size buffer where older undelivered packets are replaced by the fresher ones, we can model the dynamics of packet updates via a Geo/G/1/2 queue with replacement, as illustrated in Fig.~1.  
Because the network is considered to be static, we condition on the node positions $\Phi \triangleq \tilde{\Phi} \cup \bar{\Phi}$ and define the conditional transmission success probability of the typical link at time slot $t$ as follows \cite{YanQue:19}
\begin{align}\label{equ:CndTX_Prob}
\mu^\Phi_{0,t} = \mathbb{P}\big(\mathrm{SINR}_{0,t} > \theta | \Phi\big)
\end{align}
where $\theta$ is the decoding threshold.

Due to the broadcast nature of wireless medium, transmissions over the link pairs are correlated such that the status of any given queue is dependent on the status of the other queues and their packet depletion rates.
This phenomenon is commonly known as the spatially interacting queues \cite{SanBac:17,SanBacFos:19}, which results in $\{\mu^\Phi_{j,t}\}_{j \in \mathbb{N}, t \geq 0 }$ being correlated over space and time.
Assessing the performance of large scale wireless networks by taking into account the effect of space-time queueing interactions is a notoriously hard problem where no comprehensive theory is available at this stage.
Fortunately, when the nodes are massively deployed in space, the temporal correlations amongst their buffer states become insignificant \cite{ZhoMaoGe:20JSAC}. 
In that respect, we adopt the following approximation for tractability.
\begin{assumption}
\textit{Each node experiences independent interferers over time, and hence their queues evolve independently from each other.}
\end{assumption}
This assumption is usually referred to as the \textit{mean-field approximation} \cite{BorMcDPro:10}, which allows one to represent a varying environment by its time-average state.
Consequently, the evolution of each queue can be isolated from the current state of the network, while the effect of the spatial interactions is captured through the time-average. Notably, when the number of transmitter-receiver pairs approaches infinity the mean-field approximation has been shown to be exact asymptotically \cite{BorMcDPro:12} and hence can be applicable to the spatiotemporal analysis of large-scale networks \cite{ChiElSCon:19}.

\subsubsection{Conditional Age of Information}
Following Assumption~1, when we condition on the network topology $\Phi$, the transmissions of packets over a typical link are i.i.d. over time with a success probability $\mu^\Phi_0 = \lim_{t \rightarrow \infty} \mu^\Phi_{0,t}$.
As such, the interval of packet departures at any given link can also be approximated by a geometric distribution. In consequence, we can treat the dynamics at the typical sender as a Geo/Geo/1/2 queue where the arrival and departure rates are given by $\xi$ and $p \mu^\Phi_{0}$, respectively. 
In consequence, a conditional form of the AoI can be derived by leveraging tools from queueing theory. 
Although this result has been derived in the existing literature, we state it in the following lemma for the sake of completeness.
\begin{lemma} \label{lma:Cndt_AoI}
\textit{
	Conditioned on the point process $\Phi$, the average and peak AoI at the typical link are given respectively as follows:
	\begin{align} \label{equ:Cnd_AoI_LCFS}
	\mathbb{E}^0\!\big[\, \bar{\Delta}_0 \vert \Phi \,\big] \!&= \frac{1}{\xi} \!+\! \frac{1}{ p \mu^\Phi_0 } - 1, \\ \label{equ:Cnd_PAoI_LCFS}
	\mathbb{E}^0\!\big[\, \hat{\Delta}_0 \vert \Phi \,\big] \!&= \frac{ 1 }{\xi} \!+\! \frac{1}{ p \mu^\Phi_0 } + \frac{1}{1-(1 \!-\! \xi)(1 \!-\! p \mu^\Phi_0 )} - 2.
	\end{align}
}
\end{lemma}
\begin{IEEEproof}
See Appendix~\ref{apn:Cndt_AoI}.
\end{IEEEproof}

In view of Lemma~\ref{lma:Cndt_AoI}, we note that the core of analyzing the AoI lies at the characterization of the transmission success probability. In the following, we detail the procedure of deriving this quantity.

\subsection{ Transmission Success Probability }
Using Assumption~1, the packet departure processes at the wireless links can be assumed to be independent from each other, which indicates each node activates independently in the steady state. Then, using a similar approach as \cite{YanQue:19}, we can compute the conditional transmission success probability as follows.
\begin{lemma} \label{lma:Cnd_SucProb}
\textit{Conditioned on the network topology $\Phi$, the probability of successful transmission over the typical link is given as:
\begin{align}\label{equ:mu_0Phi}
\mu^\Phi_0 = e^{-\frac{ \theta r^\alpha}{\rho} } \prod_{ j \neq 0 } \Big( 1 - \frac{ p a^\Phi_j }{ 1 + \Vert X_j \Vert^\alpha / \theta r^\alpha } \Big)
\end{align}
where $\rho = P_{\mathrm{tx}}/\sigma^2$ is the signal-to-noise ratio (SNR) and $a^\Phi_j = \lim_{ T \rightarrow \infty} \sum_{t=0}^{T} \zeta_{j,t}/T$ the buffer non-empty probability of node $j$ in the steady state.
}
\end{lemma}
\begin{IEEEproof}
See Appendix~\ref{apn:Cnd_SucProb}.
\end{IEEEproof}

We can now explicitly identify the randomness in the departure rate, which mainly arises from $i$) the random locations of the other transmitters, and $ii$) their buffer  states. A conditional expression for the state of having a non-empty buffer at each transmitter can be obtained as follows.

\begin{lemma} \label{lma:Cndt_ActProb}
\textit{Conditioned on the network topology $\Phi$, the buffer non-empty probability of a generic node $j$ is given as:
\begin{align}\label{equ:aj}
a^\Phi_j = \frac{ \xi }{ \xi + ( 1 - \xi ) \, p \, \mu^\Phi_j  },
\end{align}
where $\mu^\Phi_j$ denotes the conditional transmission success probability of link $j$.
}
\end{lemma}
\begin{IEEEproof}
See Appendix~\ref{apn:Cndt_ActProb}.
\end{IEEEproof}

\setcounter{equation}{\value{equation}}
\setcounter{equation}{10}
\begin{figure*}[t!]
\begin{align} \label{equ:Meta_Grl}
F(u) = \frac{1}{2} -\! \int_{0}^{\infty} \!\!\! \mathrm{Im}\bigg\{ u^{-j\omega} \exp\!\Big(\! - \frac{ j \omega \theta r^\alpha }{ \rho } -  \lambda \pi r^2 \theta^\delta \sum_{k=1}^{\infty} \binom{j \omega}{ k } \!\int_0^\infty \frac{ (-1)^{k+1} dv }{ (1+v^{ \frac{ \alpha }{ 2 } } )^k }  \int_{0}^1 \!\! \frac{ (p \xi)^k d F(t) }{ [\, \xi + (1-\xi) p t \,]^k } \Big) \bigg\} \frac{ d \omega }{ \pi \omega }
\end{align}
\setcounter{equation}{\value{equation}}{}
\setcounter{equation}{11}
\centering \rule[0pt]{18cm}{0.3pt}
\end{figure*}
\setcounter{equation}{11}


With these results in hand, we can now put the pieces together and derive the distribution of the conditional transmission success probability using a similar method as \cite{YanQue:19}.

\begin{theorem} \label{thm:Meta_SINR}
  \textit{
  The cumulative distribution function (CDF) of the conditional transmission success probability is given by the fixed-point equation \eqref{equ:Meta_Grl} at the top of this page, in which $j=\sqrt{-1}$ and $\mathrm{Im}\{ \cdot \}$ denotes the imaginary part of a complex quantity.
  }
\end{theorem}
\begin{IEEEproof}
See Appendix~\ref{apn:Meta_SINR}.
\end{IEEEproof}

Owing to the space-time coupling amongst the queues, the transmission success probability CDF \eqref{equ:Meta_Grl} is given in the form of a fixed-point functional equation. It is noteworthy that the right hand side of \eqref{equ:Meta_Grl} constitutes a contraction as a functional of $F(\cdot)$. As such, solution of \eqref{equ:Meta_Grl} can be obtained via successive approximations \cite{YanQue:19}, e.g., the Picard's method, which converges exponentially fast.
Nevertheless, in each round of iteration, calculating the right hand side of \eqref{equ:Meta_Grl} requires full knowledge of all the moments of $\mu^\Phi_0$, which may be computationally troublesome. For that reason, we opt for an approximation to accelerate the calculation.

\begin{corollary} \label{cor:beta_approx}
\textit{The probability density function (PDF) of $F(u)$ in Theorem~\ref{thm:Meta_SINR} can be tightly approximated by the following:
\begin{align} \label{equ:fX}
f(u) &= \lim_{ n \rightarrow \infty } f_n(u)
\nonumber\\
&= \lim_{ n \rightarrow \infty } \frac{ u^{ \frac{ \kappa_n ( \beta_n + 1 ) - 1 }{ 1 - \kappa_n } } (1-u)^{ \beta_n - 1 } }{ B( \kappa_n \beta_n / (1-\kappa_n), \beta_n ) }
\end{align}
where $B(\cdot, \cdot)$ denotes the Beta function \cite{AndAsk:00}, $\kappa_n$ and $\beta_n$ are respectively given as:
\begin{align} \label{equ:mu_n}
\kappa_n &= c_n^{(1)}, \\ \label{equ:beta_n}
\beta_n &= \frac{ ( 1 - \kappa_n ) \big[\kappa_n - c_n^{(2)} \big] }{c_n^{(1)} - \kappa_n^2 }
\end{align}
where $c_n^{(m)}$ takes the following form:
\begin{align} \label{equ:Momnt_Beta}
c_n^{(m)} \!=\! \exp\!\bigg(\!\! -\! \frac{ m \theta r^\alpha }{ \rho }  \! - \! \lambda \pi r^2 \theta^\delta \! \sum_{k=1}^{m} \! \binom{ m }{ k }  \, {\eta}^{(k)}_n \!\bigg),
\end{align}
in which ${\eta}^{(k)}_n$ is given by
\begin{align}
{\eta}^{(k)}_{n-1} & =\int_{0}^\infty \! \frac{ (-1)^{ k+1 } dv }{ (1+v^{ \frac{\alpha}{2} })^k } \int_0^1 \!\! \frac{ (p \xi)^k f_n (t) dt }{ [\, \xi + (1-\xi) p t \,]^k }.
\end{align}
Particularly, when $n=1$, we have ${\eta}_{0}^{(k)}$ given by the following
\begin{align}
{\eta}_{0}^{(k)} = (-1)^{ k+1 } \binom{ \, \delta - 1 \,}{\, k - 1 \,} \frac{ 2 \pi^2 \theta^{ \delta } \xi^k p^k }{ \alpha \sin(  \pi \delta ) }.
\end{align}
 }
\end{corollary}
\begin{IEEEproof}
See Appendix~\ref{apn:beta_approx}.
\end{IEEEproof}

Following the above result, in each iteration, it only updates the approximation of the first and second moments of the random variable $\mu^\Phi_0$. Therefore, the procedure per Corollary~\ref{cor:beta_approx} can be carried out very efficiently.\footnote{As demonstrated in \cite{ZhaYanShe:20}, these types of recursive calculations converge in a few, e.g., less than 10, iterations. }

\subsection{Average and Peak AoI}
We are now ready to present the main results of this paper, i.e., the analytical expressions for the AoI.

\subsubsection{Average AoI}
We first present the average AoI of the network.
\begin{theorem} \label{thm:AoI_LCFS}
\textit{
	The network average AoI is given as follows:{\footnote{Note that the integral may be unbounded under certain settings of network parameters due to the singularity at the origin, which implies the interference is excessively strong. Fortunately, such a limit exists for most practical cases.}}
	\begin{align} \label{equ:EctForm_AoI_LCFS}
	\bar{ \Delta } &= \frac{1}{\xi} + \!\! \int_{0}^{1} \!\! \frac{ F(dt) }{ p t } -1 \\
    &\approx \frac{1}{\xi} +  \!\! \int_{0}^{1} \!\! \frac{  f(t) dt }{ p t } -1,
	\end{align}
    where $F(\cdot)$ and $f(\cdot)$ are given by \eqref{equ:Meta_Grl} and \eqref{equ:fX}, respectively.
}
\end{theorem}
\begin{IEEEproof}
By deconditioning \eqref{equ:Cnd_AoI_LCFS} according to the CDF and PDF of $\mu^\Phi_0$ per \eqref{equ:Meta_Grl} and \eqref{equ:fX}, respectively, the result follows.
\end{IEEEproof}

Notably, the AoI expressions in Theorem~\ref{thm:AoI_LCFS} account for several key features of a random access network, including the packet arrival rate, channel access probability, deployment density, and interference.
We will verify the accuracy of this analysis in Section~IV and obtain a number of design insights based on numerical results.
Before that, let us focus on two regimes of operation to develop a deeper understanding of the network average AoI.

\remark{\textit{ When $\lambda \rightarrow 0$, the network is in the noise-limited regime, i.e., the SINR expression in (5) becomes
\begin{align}
\mathrm{SINR}_0 \approx \frac{P_{\mathrm{tx}} H_{00} r^{-\alpha} }{ \sigma^2 }.
\end{align}
Then, by jointly using (6) and (7), it can be shown that the network average AoI is given by
\begin{align}
\bar{\Delta} = \frac{1}{\xi} - 1 + \frac{\exp\big( \frac{\theta r^\alpha}{\rho} \big)}{p},
\end{align}
which monotonically decreases with the packet arrival rate $\xi$.
}}
This observation is in line with conclusions drawn from the conventional point-to-point settings, namely under the LCFS discipline, increasing the update frequency can always benefit the AoI performance.

We next investigate the AoI in the interference limited regime, namely $\rho \gg 1$ and $\lambda$ is relatively large.
In lieu of directly dealing with the original system, let us consider the transmissions undergo a \textit{favorable system}, in which the incoming packets are sent out only once -- regardless of the transmission being successful or not -- without retransmissions. We denote the conditional transmission success probability achieved at the typical link in such a system as $\check{\mu}^\Phi_0$. 
Because every node in the favorable system only activates when a new packet arrives, the buffer non-empty probability of a generic node $j$ is $\check{a}^\Phi_j = \xi$, which satisfies $\check{a}^\Phi_j \leq a^\Phi_j$ according to \eqref{equ:aj}. Then, following \eqref{equ:mu_0Phi} we know that transmissions in a favorable system suffer less interference than the original one, which yields $\mu^\Phi_0 \leq  \check{\mu}^\Phi_0$ and hence the following relationship holds
\begin{align}
 \mathbb{E}^0\big[ \bar{\Delta} | \Phi \big] =  \frac{1}{\xi} - 1 + \frac{1}{p \mu^\Phi_0} \geq \frac{1}{\xi} - 1 + \frac{1}{p \check{\mu}^\Phi_0 }.
\end{align}
As such, if we take an expectation on both sides of the above inequality, it yields
\begin{align} \label{equ:AveAoI_LU_relation}
\bar{\Delta} &\geq \frac{1}{\xi} - 1 + \mathbb{E}\Big[ \frac{1}{p \check{\mu}^\Phi_0 } \Big]
\nonumber\\
&\stackrel{(a)}{ \geq } \frac{1}{\xi} - 1 +  \frac{1}{p \mathbb{E}[\check{\mu}^\Phi_0 ] }
\nonumber\\
&\stackrel{(b)}{ = } \frac{1}{\xi} - 1 +  \frac{ \exp\Big( \lambda \pi r^2 \theta^\delta \!\! \int_0^\infty\!\!\!\! \frac{dv}{ 1 + v^{ \frac{\alpha}{2} } } \times p \, \xi \Big) }{ p } = Z(\xi, p)
\end{align}
where ($a$) follows from the Jensen's inequality, and ($b$) is by noticing that $\check{\mu}^\Phi_0$ can be obtained by assigning $a^\Phi_j = \xi$ in \eqref{equ:CndTX_Prob} and further leveraging the probability generating functional (PGFL) of PPP to carry out the calculation.

From the expression of $Z(\xi,p)$ in \eqref{equ:AveAoI_LU_relation}, it is clear that as $\xi$ increases, the inter-arrival interval time of information packets decreases hyperbolically, while the packet departure time grows exponentially. In that respect, if the gain of update freshness at the source node cannot compensate the loss in the transmission delay, increasing the packet arrival rate may not benefit the AoI.
To formally demonstrate such an observation, let us take a derivative of $Z(\xi, p)$ with respect to $\xi$ and assign it to be zero, which results in the following
\begin{align} \label{equ:LB_Derivative}
&- \frac{1}{\xi^2} +\!\!  \int_0^\infty\!\! \frac{ \lambda \pi r^2 \theta^\delta dv }{ 1 + v^{ \frac{\alpha}{2} } } \cdot \exp\Big( \lambda \pi r^2 \theta^\delta \!\!\! \int_0^\infty\!\!\!\! \frac{dv}{ 1 \!+\! v^{ \frac{\alpha}{2} } } \times p \, \xi \Big) 
\nonumber\\
&= \frac{ \partial Z(\xi,p) }{\partial \xi } = 0.
\end{align}
From the above expression, we can see that $\frac{ \partial Z(\xi,p) }{\partial \xi }$ monotonically increases with respect to $\xi$. Because $\xi \in (0,1]$,  and $\frac{ \partial Z(\xi,p) }{\partial \xi } \rightarrow - \infty$ as $\xi \rightarrow 0$, it is clear that \eqref{equ:LB_Derivative} possesses a unique solution if $\frac{ \partial Z(\xi,p) }{\partial \xi } |_{\xi = 1} > 0$, which is equivalent to the following condition
\begin{align}
\lambda > \lambda_0 = \frac{ \mathcal{W}_0(p) }{ p \pi r^2 \theta^\delta \!\! \int_{0}^{\infty} \! \frac{ dv }{ 1 + v^{ \alpha / 2 } }  },
\end{align}
where $\mathcal{W}_0(\cdot)$ is the Lambert function. Otherwise, we have $\frac{ \partial Z(\xi,p) }{\partial \xi } < 0$ for all $\xi \in (0, 1]$, namely $Z(\xi, p)$ always decreases with $\xi$.
That motivates us to make the following remark.

\remark{\textit{ Given distance $r$, there exists a threshold of the deployment density $\lambda_0$, when $\lambda > \lambda_0$, the average AoI does not monotonically decrease with the packet arrival rate $\xi$.
}}

Similarly, by assigning $\frac{\partial Z(\xi, p)}{\partial p} = 0$, we have the following:
\begin{align}
&- \frac{1}{p^2} \exp\Big( \lambda \pi r^2 \theta^\delta \!\! \int_0^\infty\!\!\!\! \frac{dv}{ 1 + v^{ \frac{\alpha}{2} } } \times p \, \xi \Big)
\nonumber\\
& + \frac{\lambda \pi r^2 \theta^\delta \!\! \int_0^\infty\!\!\!\! \frac{  \xi  dv}{ 1 + v^{ \frac{\alpha}{2} } } }{p} \exp\Big( \lambda \pi r^2 \theta^\delta \!\! \int_0^\infty\!\!\!\! \frac{dv}{ 1 + v^{ \frac{\alpha}{2} } } \times p \, \xi \Big) = 0.
\end{align}
The solution to the above leads us to the following remark. 
\remark{\textit{ In a random access network, the optimal $p$ that minimizes the average AoI is approximately given by
\begin{align}
p^* = \min\bigg\{ 1, \frac{1}{ \xi \lambda \int_{0}^{\infty} \! \frac{ \pi r^2 \theta^\delta dv }{ 1 + v^{\alpha/2} } } \bigg\}.
\end{align}
}}

Albeit the above insights are drawn from the lower bound $Z(\xi, p)$ obtained from a favorable system, they can be interpreted as a simple proxy to the original system and are useful for the understanding of effects from deployment and interplay between spatial and temporal factors of a random access network on the AoI. As we will see in the section of numerical results, similar observations to the network average AoI $\bar{\Delta}$ occur in the original system.

\subsubsection{Outage probability of peak AoI}
Next, we look at the outage probability of peak AoI, defined as the probability that the peak AoI of any given link exceeds a threshold $A$.
The reason for adopting such a metric is that it is relevant to system designs that guarantee the information available at the destination is fresh at any given time \cite{CosCodEph:16}, while an average value of the peak AoI can also be derived on similar lines of Theorem~2.
\begin{theorem}
\textit{
	The outage probability of peak AoI is given by:
\begin{align}
& \mathbb{P}\big( \mathbb{E}^0[ \hat{\Delta}_0 | \Phi] > A \big)
\nonumber\\
& = F\left( \frac{ - \xi (1+c) + \sqrt{ \xi^2 c^2 + 4 \xi c + 2 \xi^2 c - 3 \xi^2 } }{ 2 c (1-\xi) p } \right)
\end{align}
where $F(\cdot)$ is given in \eqref{equ:Meta_Grl} and $c$ is given by
\begin{align}
c = A + 2 - \frac{1}{\xi}.
\end{align}
}
\end{theorem}
\begin{IEEEproof}
Using the expression for conditional peak AoI in \eqref{equ:Cnd_PAoI_LCFS}, the event $\{ \mathbb{E}^0[ \tilde{\Delta}_0 | \Phi ] > A \}$ can be expanded as follows
\begin{align}
\frac{1}{ p \mu^\Phi_0 } - \frac{1}{ 1 - (1-\xi)( 1 - p \mu^\Phi_0 ) } > A + 2 - \frac{1}{\xi} = c.
\end{align}
By rearranging the terms, we have the following
\begin{align}
c(1-\xi) (p \mu^\Phi_0)^2 + \xi c p \mu^\Phi_0 + \xi p \mu^\Phi_0 - \xi < 0.
\end{align}
Solving the above inequality yields
\begin{align}
0< \mu^\Phi_0< \frac{ - \xi (1+c) + \sqrt{ [\xi (1+c)]^2 + 4 \xi c (1-\xi) } }{ 2 c (1-\xi) p }.
\end{align}
The result then follows by deconditioning $\mu^\Phi_0$ in the above inequality using \eqref{equ:Meta_Grl} along with further algebraic manipulations.
\end{IEEEproof}

\section{Locally Adaptive Slotted ALOHA for AoI Minimization }
It has been shown in \cite{YanArafaQue:19} that when availed with local information of the network geometry at each node, a locally adaptive channel access scheme can be devised to reduce the network peak AoI under the FCFS discipline. A natural question then arises as: \textit{ Can we develop a similar approach to minimize the AoI in random networks where the transmitters are equipped with unit-size buffers and sending out packets under a LCFS-R protocol?}
We give an affirmative answer to this question in this section.

\subsection{Stopping Sets and Objective Function}
\begin{figure}[t!]
  \centering{}

    {\includegraphics[width=0.95\columnwidth]{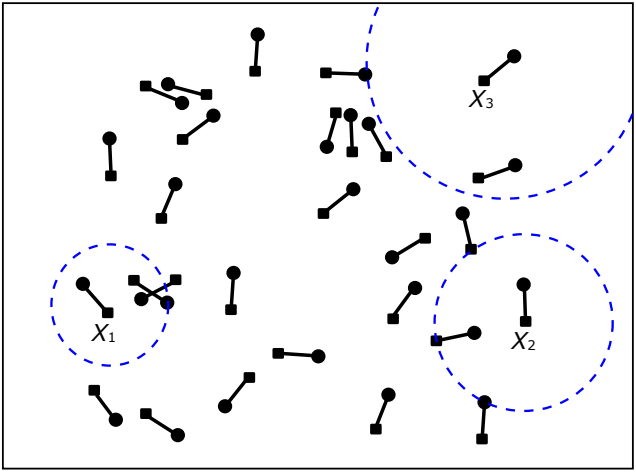}}

  \caption{ Example of a Poisson network in which every transmitter can observe the closest and second closest receivers to it. Here, the black squares and dots are the transmitters and receivers, respectively, and the circles with dashed lines are stopping sets centered at $X_1$, $X_2$, and $X_3$. }
  \label{fig:StpSet_V1}
\end{figure}

Our objective is to derive a link-wise channel access probability that minimizes the network average AoI based on the local information observed from each individual node.
Due to the limited sensing capability, every transmitter can only obtain knowledge about a finite region around it, which is denoted by the observation window $W$.
Such an observation window generally possesses a random shape owing to the various sensing capability of transmitters. 
In that respect, we leverage the concept of stopping sets \cite{BacBla:09,BacBlaSin:14} to describe the arbitrary shape of the observation window.
Specifically, a stopping set is a random element takes the form in Borel sets such that for any observation window $\mathcal{O}$, one can determine whether $W = W(\tilde{\Phi}, \bar{\Phi}) \subset \mathcal{O}$ is true or not.
Depending on the scenarios under consideration, stopping sets can take different forms.
An example of random stopping sets in a Poisson bipolar network is given in Fig.~\ref{fig:StpSet_V1}. Note that the stopping sets associated with different transmitters, e.g., the nodes located at $X_1$, $X_2$, and $X_3$, can have various shapes.

In consequence, the channel access probability constructed at the typical node takes the following form
\begin{align} \label{equ:eta_S0}
\gamma^\Phi_0 &= \eta_{_\mathrm{W}}(\tilde{\Phi}, \bar{\Phi})
\nonumber\\
&= \eta_{_\mathrm{W}}\!\left( \tilde{\Phi} \cap W, \bar{\Phi} \cap W \right),
\end{align}
where $\eta_{_{W}}(\cdot) \in [0, 1]$ is a measurable function whose argument is the network information, i.e., the buffer states as well as locations of the nodes in ($\tilde{\Phi} \cap W$, $\bar{\Phi} \cap W$).
For a node $i$ located at $X_i$, its scheduling policy can be obtained by applying the shifting operator $S_{X_i}$ to \eqref{equ:eta_S0}, which moves the origin of point process $\Phi$ to $X_i$ and results in
\begin{align} \label{equ:etaW_Gnl}
\gamma^\Phi_i &= S_{X_i} \eta_{_W}
\nonumber\\
&= \eta_{_W}\left( S_{X_i} \tilde{\Phi} \cap W, S_{X_i} \bar{\Phi} \cap W \right).
\end{align}
To this end, the design of our channel access policy can be cast into the following optimization problem:
\begin{subequations} \label{prbm:minAoI}
\begin{align} \label{prbm:Dom_PeakAoI}
& \min_{ \eta_{_\mathrm{W}} } ~~~~ \mathbb{E}^0_\Phi\! \left[ \frac{ 1 }{ \gamma_0^\Phi {\mu}_0^{\Phi | W} } \right] + \frac{1}{\xi} - 1 \\ \label{equ:Dom_Rst_SpSt}
& ~~ \mathrm{s.t.} \quad~ 0 \leq \gamma_i^\Phi = S_{X_i} \eta_{_\mathrm{W}} \leq 1, ~~ \forall~ i
\end{align}
\end{subequations}
where ${\mu}_0^{\Phi | W} = \mathbb{E}^0_\Phi\big[ \mu^\Phi_0 | W \big]$ is the conditional transmission success probability given observation window $W$, and \eqref{equ:Dom_Rst_SpSt} stipulates the channel access probability devised at each individual node to be within a feasible region.
It is worthwhile to note that ($a$) the local information varies across the nodes and so as their channel access probabilities, and ($b$) such a policy can be carried out without the coordination of a central controller and hence is decentralized.\footnote{Note that if $W$ is set to be the whole space, then the scheduling algorithm becomes centralized, although this is not practical due to the excessive signaling overhead.}

\subsection{ Design of the Channel Access Policy }
In order to solve \eqref{prbm:minAoI}, we need to first have an expression for the quantity ${\mu}_0^{\Phi | W}$, which is given by the following.

\begin{lemma}
\textit{ Given the observation window $W$ and channel access probability $\eta_{ _\mathrm{W} }$, the conditional transmission success probability at a generic link $i$ is given as
\begin{align} \label{equ:CndTXSucProb_Domint}
{\mu}^{\Phi | W}_i = e^{ - \frac{ \theta r^\alpha }{ \rho } } \!\!\!\!\!\!\!\! \prod_{ j \neq 0, j \in W } \!\!\!\!\!\! \big( 1 - \frac{\gamma^\Phi_j a^\Phi_j }{ 1 \!+\! \mathcal{D}_{ji} } \big) \exp\Big( - \!\!\!\!\!\!\!\! \int\limits_{ \mathbf{x} \in \mathbb{R}^2 \setminus W } \!\!\!\!\!\!\!\! \frac{\gamma_{ \mathbf{x} } \mathbb{E}\big[ a^\Phi_{\mathbf{x}} \big]  d\mathbf{x} }{ 1 \!+\! \Vert \mathbf{x} \Vert^\alpha \!/ \theta r^\alpha } \Big).
\end{align}
where $\mathcal{D}_{ji} = \Vert X_j - y_i \Vert^\alpha/\theta r^\alpha$.
}
\end{lemma}
\begin{IEEEproof}
When every transmitter adopts the channel access policy $\eta_{ _\mathrm{W} }$ per \eqref{equ:etaW_Gnl}, given the observation window $W$ and using Assumption~1, we can use similar approaches in the derivation of Lemma~\ref{lma:Cnd_SucProb} and arrive at the following
\begin{align}
{\mu}^{\Phi | W}_i &= e^{-\frac{ \theta r^\alpha}{\rho} } \cdot \mathbb{E}^{X_i}_\Phi\bigg[ \prod_{ j \neq i, j \in W } \Big( 1 - \frac{ \gamma^\Phi_j a^\Phi_j }{ 1 + \mathcal{D}_{ji} } \Big) \Big\vert W \bigg]
\nonumber\\
&\qquad \qquad \times \mathbb{E}^{X_i}_\Phi \bigg[ \prod_{ j \notin W } \Big( 1 - \frac{ \gamma^\Phi_j a^\Phi_j }{ 1 + \mathcal{D}_{ji} } \Big) \Big\vert W \bigg]
\nonumber\\
&\stackrel{(a)}{=} e^{-\frac{ \theta r^\alpha}{\rho} } \cdot  \prod_{ j \neq i, j \in W } \Big( 1 - \frac{ \gamma^\Phi_j a^\Phi_j }{ 1 + \mathcal{D}_{ji} } \Big)
\nonumber\\
&\qquad \qquad \times \mathbb{E}^{X_i}_\Phi \bigg[ \prod_{ j \notin W } \Big( 1 - \frac{ \gamma^\Phi_j a^\Phi_j }{ 1 + \mathcal{D}_{ji} } \Big) \Big\vert W \bigg]
\end{align}
where ($a$) follows by the tower property of conditional probability, and the final result can then be derived by using the PGFL of PPP for further calculation.
\end{IEEEproof}

Using the above result, we can now solve \eqref{prbm:minAoI} as follows.
\begin{theorem}\label{thm:DmSym_PeakAoI}\textit{For all given stopping sets $W = W(\tilde{\Phi}, \bar{\Phi})$, the solution to the optimization problem in \eqref{prbm:minAoI} is given by the unique solution of the following fixed point equation:
	\begin{align}\label{equ:OptSln}
	\frac{1}{\eta_{_\mathrm{W}} } -\!\!\!\!\! \sum_{ \substack{ j \neq 0, y_j \in W } } \frac{1}{ 1  \!+\! \mathcal{D}_{0j} \!-\! a^\Phi_j \eta_{_\mathrm{W}} } -\!\! \int_{\mathbb{R}^2 \setminus W }\! \frac{ \lambda \mathbb{E}\big[ a^\Phi_0 \big] dz }{ 1 \!+\! \Vert z \Vert^\alpha \! / \theta r^\alpha } = 0
	\end{align}
	if the following condition holds
	\begin{align} \label{equ:CndOptl}
	\sum_{ \substack{ j \neq 0, y_j \in W } } \frac{1}{ 1  \!+\! \mathcal{D}_{0j} \!-\! a^\Phi_j } +\! \int_{ \mathbb{R}^2 \setminus W }\! \frac{ \lambda \mathbb{E}\big[ a^\Phi_0 \big]  dz }{ 1 \!+\! \Vert z \Vert^\alpha \! / \theta r^\alpha  } > 1.
	\end{align}
	Otherwise, $\eta_{_\mathrm{W}} = 1$.
}
\end{theorem}
\begin{IEEEproof}
See Appendix~\ref{apx:OptCtrl_PeakAoI}.
\end{IEEEproof}

It is important to emphasize that $\eta_{_W}$ is in essence a function that takes in local information and produces a channel access probability, where the above theorem only presents an example of constructing the channel access policy at the typical link, namely $\gamma_0^\Phi = \eta_{ _{W} }(S_{ X_0 } \tilde{\Phi}, S_{ X_0 } \bar{\Phi} )$. In our employed network, different transmitters can have disparate local observations and hence will generate different channel access probabilities. Specifically, for a generic node $i$, the corresponding policy can be attained by applying a shifting operator $S_{X_i}$ to Theorem~\ref{thm:DmSym_PeakAoI}, i.e., by moving the origin of the network to $X_i$, which results in $\gamma_i^\Phi = \eta_{ _{W} }(S_{ X_i } \tilde{\Phi}, S_{ X_i } \bar{\Phi} )$. As such, every node in this network only needs to identify and record the transmitting neighbors located inside its observation window, i.e., the stopping set $W$, and solve for the channel access probability via a fixed point equation.

\begin{algorithm}[t!]
\caption{ Locally Adaptive Slotted ALOHA }
\begin{algorithmic}[1] \label{agl:Sptmp_Schdul}
\State \textbf{Parameters:} $\gamma^\Phi_{i,t}$: Channel access probability of link $i$ at time slot $t$, $a^\Phi_{i,t}$: Buffer non-empty probability of link $i$ at time slot $t$
\State \textbf{Initialize:} Set $\gamma^\Phi_{j,0} = 1$, $\forall j \in \mathbb{N}$, transmitters update the location
 information with neighbors inside $W = W(\tilde{\Phi}, \bar{\Phi})$
\For { time slot $t$ }
\If  {$t \neq 0 \, ( \mathrm{mod}~ T_m ) $}
\State  $\forall j \in \mathbb{N}$, make the channel access decision according to $\gamma^\Phi_{j,t}$, and record the corresponding buffer non-empty probability $a^\Phi_{j,t}$
 \Else
 \State For each link $i$, updates $a^\Phi_{i,t}$ to, and also receives $a^\Phi_{j,t}$ from, all the $X_j \in S, j \neq i$, recalculate the value of $\gamma^\Phi_{i,t}$ according to Theorem~\ref{thm:DmSym_PeakAoI}
\EndIf
\State $t \gets t + 1 $
\EndFor
\end{algorithmic}
\end{algorithm}

According to Theorem~\ref{thm:DmSym_PeakAoI}, the implementation of the proposed policy requires transmitters to monitor their queue status and mutually update
the information about their buffers' state. In particular, each transmitter needs to first identify and record the transmitting neighbors that are located inside the stopping set $W$  \cite{BacBlaSin:14}. Additionally, every transmitter will also need to collect the updates about the buffer non-empty probabilities from the neighboring transmitters \cite{KimDeYan:12}.
However, updating the local information every time slot can incur a hefty amount of overhead that degrades the efficiency. To overcome this problem, we reduce the update frequency of local information by combining $m$ consecutive time slots into a frame, denoted as $T_m$, and the updates of mutual information only take place at the beginning of each time frame.
The entire implementation process is summarized in Algorithm~1.
It shall be also noted that the solution given in this paper is based on the mean-field approximation as no comprehensive theory on the exact characterization of the original tystem is available at this stage. 
Moreover, the scheme differs from the previous result \cite{YanArafaQue:19} in that it accounts for the buffer non-empty probability in the policy design. We will show in Section V that such a subtle change can actually lead to a significant difference in the performance of the algorithm.

\remark{\textit{From Theorem~\ref{thm:DmSym_PeakAoI}, we can see that if $\mathcal{D}_{j0}$ decreases for some $j$, the channel access probability $\eta_{ _\mathrm{W} }$ also decreases, namely the nodes located in a crowded area of space automatically reduces their channel access frequencies to reduce interference, and vice versa.
}}

\remark{\textit{It can be shown that the channel access probability given in Theorem~\ref{thm:DmSym_PeakAoI} also minimizes the average peak AoI in the employed system. As such, the proposed scheme is applicable to reduce both the average AoI and average peak AoI.
}
}

Following similar arguments as \cite{BacBlaSin:14}, it can be shown that the proposed channel access policy is also capable of maximizing the logarithm of throughput in the considered network. In this regard, the gain in information freshness is essentially brought by enhancing the link throughput during the packet transmission phases.
Moreover, if the buffer non-empty probability $a_j$ is assigned to be one for all the nodes, i.e., $a_j = 1, \forall j \in \mathbb{N}$, in Theorem~4, the solution reduces to that proposed in \cite{YanArafaQue:19}. 
It can be shown that in networks where no ACK/NACK message is available from the receivers, and the transmitters keep sending out each information packet untill the arrival of a new one, such a policy can be employed to minimize the AoI.

\section{Simulation and Numerical Results}
In this section, we show simulation results that confirm the accuracy of our analytical framework, and based on the analysis we further investigate the AoI performance under different settings of network parameters.
Particularly, we consider a square region with side length of 1 km, in which link pairs are scattered according to a Poisson bipolar network with spatial density $\lambda$ and once the topology is generated it remains unchanged.
To eliminate the favorable interference coordinations induced by network edges, we use wrapped-around boundaries \cite{FasMueRup:19} that allow dipoles that leave the region on one side to reappear on the opposite side, thus mirroring the missing interferers beyond the scenario boundary.
Then, the dynamics of status update at each link are run over 10,000 time slots.
Specifically, at the beginning of each time slots, channel gains are independently instantiated and packets are generated at each transmitter with probability $\xi$, whereas the newly arriving packets replace the older undelivered one at each node.
The nodes with non-empty buffers then send out packets to the destination with failure retransmission occur at the next time slot (unless the packet is replaced by a newly generated one). And a packet can be dropped from the transmitter queue if the received SINR at the intended node exceeds the decoding threshold.
The AoI statistics of the receivers of all the links are recorded to construct the average AoI.
Unless differently specified, we use the following parameters: $\alpha = 3.8$, $\theta=0$~dB, $P_{\mathrm{tx}}=17$~dBm, $T_m = 200$, and $\sigma^2 = -90$~dBm.

\begin{figure}[t!]
  \centering{}

    {\includegraphics[width=0.95\columnwidth]{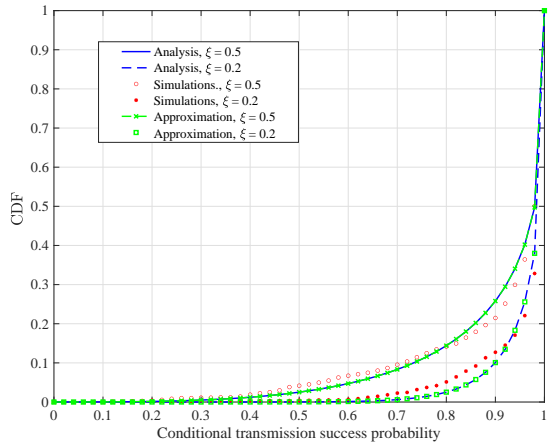}}

  \caption{ Simulation versus analysis, where we set $p=1$, $r=0.5$ m, $\lambda = 1 \times 10^{-2}$ m$^{-2}$, and vary the packet arrival rates as $\xi = 0.2, 0.5$. }
  \label{fig:VerAnaly}
\end{figure}

Fig.~\ref{fig:VerAnaly} compares the simulated CDF of the conditional transmission success probability to the analyses given in Theorem~1, as well as the approximations in Corollary~1, under different values of packet update frequency $\xi$.
From this figure, we can see that the analyses match well with simulations, which confirms the appropriateness of adopting the mean-field approximation in the analytical derivations.
Besides, the differences between the analysis in \eqref{equ:Meta_Grl} and approximation per \eqref{equ:fX} are almost indistinguishable, which verifies the tightness of approximating the meta distribution via a Beta distribution.

\begin{figure}[t!]
  \centering{}

    {\includegraphics[width=0.95\columnwidth]{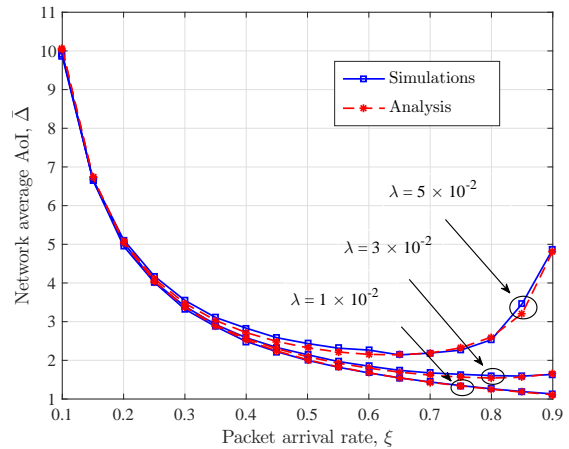}}

  \caption{ Simulation versus analysis of the network average AoI, in which we set $p=1$, $r = 0.5$ m, and vary the deployment densities as $\lambda = 1 \times 10^{-2}, 3 \times 10^{-2}, 5 \times 10^{-2}$ m$^{-2}$. }
  \label{fig:SimVerf}
\end{figure}

In Fig.~\ref{fig:SimVerf}, we depict the network average AoI as a function of the packet arrival rate $\xi$, under different values of the deployment density $\lambda$.
From this figure, we first observe a close match between the simulation and analytical results, which verifies the accuracy of Theorem~\ref{thm:AoI_LCFS}.
Moreover, we note that the optimal update frequency that minimizes the average AoI is dependent on the particular value of $\lambda$.
Specifically, when $\lambda$ is small, the link pairs recede into the distance from each other and the packet transmissions can enjoy low level of interference because of the path loss.
This resembles a noise-limited scenario and, as pointed out by Remark~1, the average AoI can be reduced by increasing the update frequency at the source nodes.
On the contrary, when $\lambda$ becomes large, the network is densely deployed, in which the inter-link distances shrink and transmitters in geographical proximity can suffer from interference that results in transmission failures.
As such, with an increase of packet arrival rate, not only more link pairs are activated but, more crucially, additional failure packet deliveries and retransmissions are incurred, which prolongs the active period of the nodes. These together slow down the packet successful decoding process at each individual link and deteriorate the information freshness over the network.
In consequence, an optimal arrival rate exists that balances the tradeoff between the information freshness at the source nodes and the interference level across the network.
This observation is consistent with Remark~2 and shows an unconventional behavior of the AoI in random access networks employing LCFS queueing disciplines.

\begin{figure}[t!]
  \centering{}

    {\includegraphics[width=0.95\columnwidth]{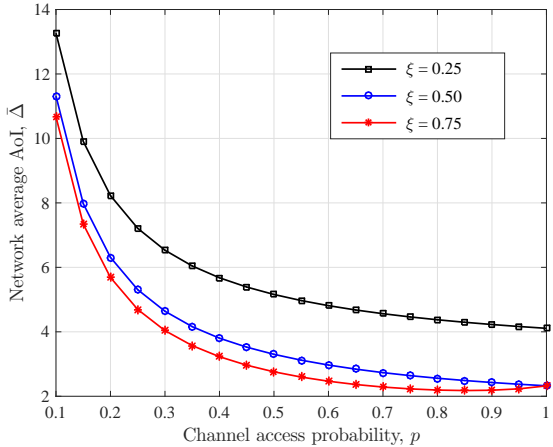}}

  \caption{ The average AoI versus channel access probability, where we set $r=0.5$ m, $\lambda = 5 \times 10^{-2}$ m$^{-2}$, and vary the packet update frequencies as $\xi = 0.25, 0.50, 0.75$. }
  \label{fig:AoI_vs_ChnlAcs}
\end{figure}

Fig.~\ref{fig:AoI_vs_ChnlAcs} plots the average network AoI for fixed $\lambda = 5 \times 10^{-2}$ as a function of the channel access probability $p$, under various packet arrival rates. We can see that in the situation of infrequent packet arrivals, the average AoI declines steadily as the channel access probability increases.
It is worth noting that this observation poses a dissent on the conclusions drawn from conflict graph models \cite{CheGatHas:19}, where the slotted ALOHA protocol is asserted to be optimal for minimizing average AoI in the light traffic condition. 
The reason for such a difference is that under the SINR model, for small update frequency, the aggregated interference at each node is low and hence there is no necessity to reduce the channel use, which will, in turn, downgrade the packet successful decoding rate and deteriorate the AoI.
Note that if the wireless links are deployed in an ultra dense manner, the SINR model converges to collion model and conclusions drawn here will be similar to those obtained in \cite{CheGatHas:19}.
On the other hand, when the packet arrival rate is high, we can see that there exists an optimal channel access probability which minimizes the average AoI. This is because the defection of interference on the service rate is more devastating in this scenario, and exerting controls on the channel access are of importance to bolster the transmissions.
As such, the slotted ALOHA protocol is beneficial to striking a balance between information freshness at the transmitters and the overall interference level.
Note that similar conclusions can be drawn from  Remark~3, showing the compliance between observations and analysis. Moreover, the figure also indicates that in order to achieve a small average AoI across the network, one should tune the update arrival frequency to a high level and adopt slotted ALOHA to control the channel access.

\begin{figure}[t!]
  \centering{}

    {\includegraphics[width=0.95\columnwidth]{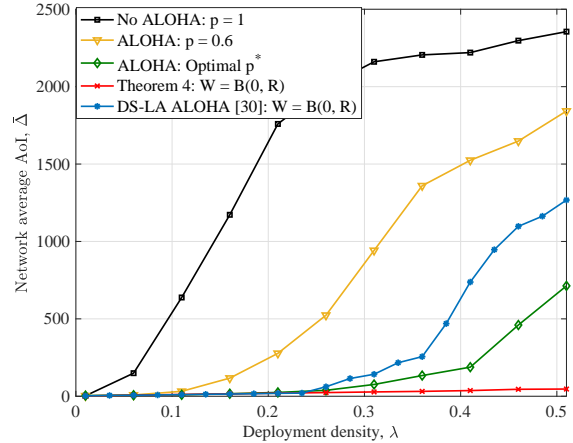}}

  \caption{ Network average AoI versus spatial deployment density, where we set $\xi=0.6$, $r = 2.5$~m, $R=20$~m, and $\lambda$ increases from $1 \times 10^{-2}$ m$^{-2}$ to $5 \times 10^{-1}$ m$^{-2}$.  }
  \label{fig:AoI_sptltmr_ctrl}
\end{figure}

In Fig.~\ref{fig:AoI_sptltmr_ctrl} we compare the proposed channel access policy per Theorem~\ref{thm:DmSym_PeakAoI} to the slotted ALOHA protocol as well as the one proposed in \cite{YanArafaQue:19}, which we termed as Dominant System-based Locally Adaptive ALOHA (DS-LA ALOHA). Specifically, we set the local observation window at each node as a deterministic disk which centered at the transmitter with a radius $R$, i.e., $W=B(0,R)$, and vary the slotted ALOHA channel access probability $p$ as 1, 0.6, and the optimal $p^*$ which is tuned according to the variants of network parameters. From this figure, we first notice that without controls on the channel access, namely $p=1$, the network average AoI soars sharply when the spatial deployment is densified. This mainly attributes to the fact that as deployment density increases, the mutual link distances shrank, and that ramps up the interference level which inflicts failures on the transmissions. 
In contrast, with the adoption of slotted ALOHA, the network immediately attains a large reduction to the average AoI which demonstrates the efficacy of such a protocol in large-scale wireless networks. 
And more prounced gains in AoI can be observed when the parameter $p$ is optimally set. 
Additionally, when each of the nodes adopt the scheme proposed in Theorem~4, we observe a remarkable gain in the network average AoI. This is because the channel access probability is link-wisely configured based on the local observations, which marshals the spectral resource more adequately and averts transmitters in geographical vicinity to transmit simultaneously.
As a result, transmissions under the proposed scheme are able to achieve a high success rate and hence the AoI can be kept at a low level in wide regimes of deployment density.
Nonetheless, we can also see that while the DS-LA ALOHA is also able to reduce the AoI, it does not even outperform the slotted ALOHA when the latter is operating on an optimally tuned parameter. 
This is because the transmitters only maintain unit-size buffers which are less likely to be saturated, while the locally adaptive ALOHA policy constructed under the dominant system tends to overestimate the interference and reduces the frequency of channel access at each node. And that leads to unnecessarily prolonged waiting time which worsens the AoI. 
To this end, it is worthwhile to emphasize that the temporal parameters in the locally adaptive slotted ALOHA shall be adequately selected to ensure it can be functioning at full power.

We now turn our attention to investigate the effects of different network parameters on the outage probability of peak AoI.

\begin{figure}[t!]
  \centering{}

    {\includegraphics[width=0.95\columnwidth]{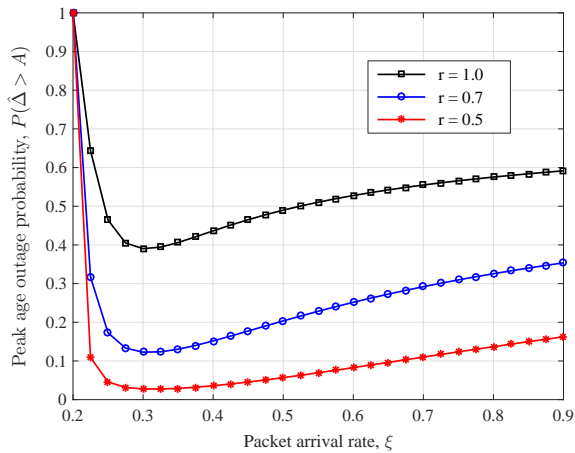}}

  \caption{ Peak age outage probability versus packet arrival rate, in which we set $p=1$, $A = 5$, $\lambda = 5 \times 10^{-2}$~m$^{-2}$ and vary transceiver distance as $r = 0.5, 0.7, 1.0$ m.  }
  \label{fig:AoI_Outage_vs_xi}
\end{figure}

Fig.~\ref{fig:AoI_Outage_vs_xi} shows the outage probability of peak AoI for a varying value of the packet arrival rate, under different distances between the transmitter-receiver pairs. We immediately notice the existence of an optimal update frequency that minimizes the outage probability, owing to a tradeoff between freshness of information from the source and the total interference level.
Note that compared to the network average AoI, in order to minimize the outage probability of peak AoI, the optimal frequency shall be set to a relatively small value. Additionally, with a slight increase of $r$, the outage probability of peak AoI sheers up rapidly, showing the peak AoI is more sensitive to the variant of network parameters.

\begin{figure}[t!]
  \centering{}

    {\includegraphics[width=0.95\columnwidth]{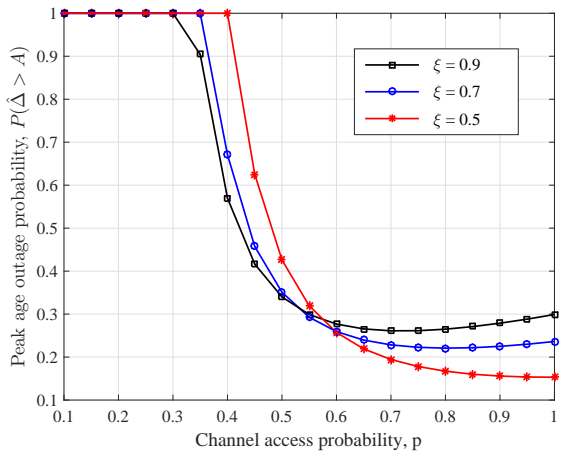}}

  \caption{ Peak age outage probability versus channel access probability, where we set $A = 5$, $r=0.7$ m, $\lambda = 5 \times 10^{-2}$ m$^{-2}$, and vary the packet arrival rates as $\xi = 0.5, 0.7, 0.9$. }
  \label{fig:AoI_Outage_vs_p}
\end{figure}

In Fig.~\ref{fig:AoI_Outage_vs_p}, we put the spotlight on the outage probability of peak AoI under different channel access probabilities.
We can see that similar to the average AoI, the outage probability of peak AoI keeps declining with respect to $p$ in the regime of low packet arrival rate, while it can be minimized by an optimal value of $p$ when the packets arrive rates are high.
Nonetheless, different from the minimization of network average AoI, to obtain a small outage probability of peak AoI, it is more desirable to reduce the packet arrival rate rather than setting it at a high level and then employ slotted ALOHA.
It is noteworthy that this observation also marks a sharp distinction to that under the FCFS discipline, which claims the slotted ALOHA protocol cannot contribute to reducing the peak AoI \cite{YanArafaQue:19}.

%
%

\begin{figure*}[t!]
  \centering

  \subfigure[\label{fig:1a}]{\includegraphics[width=0.95\columnwidth]{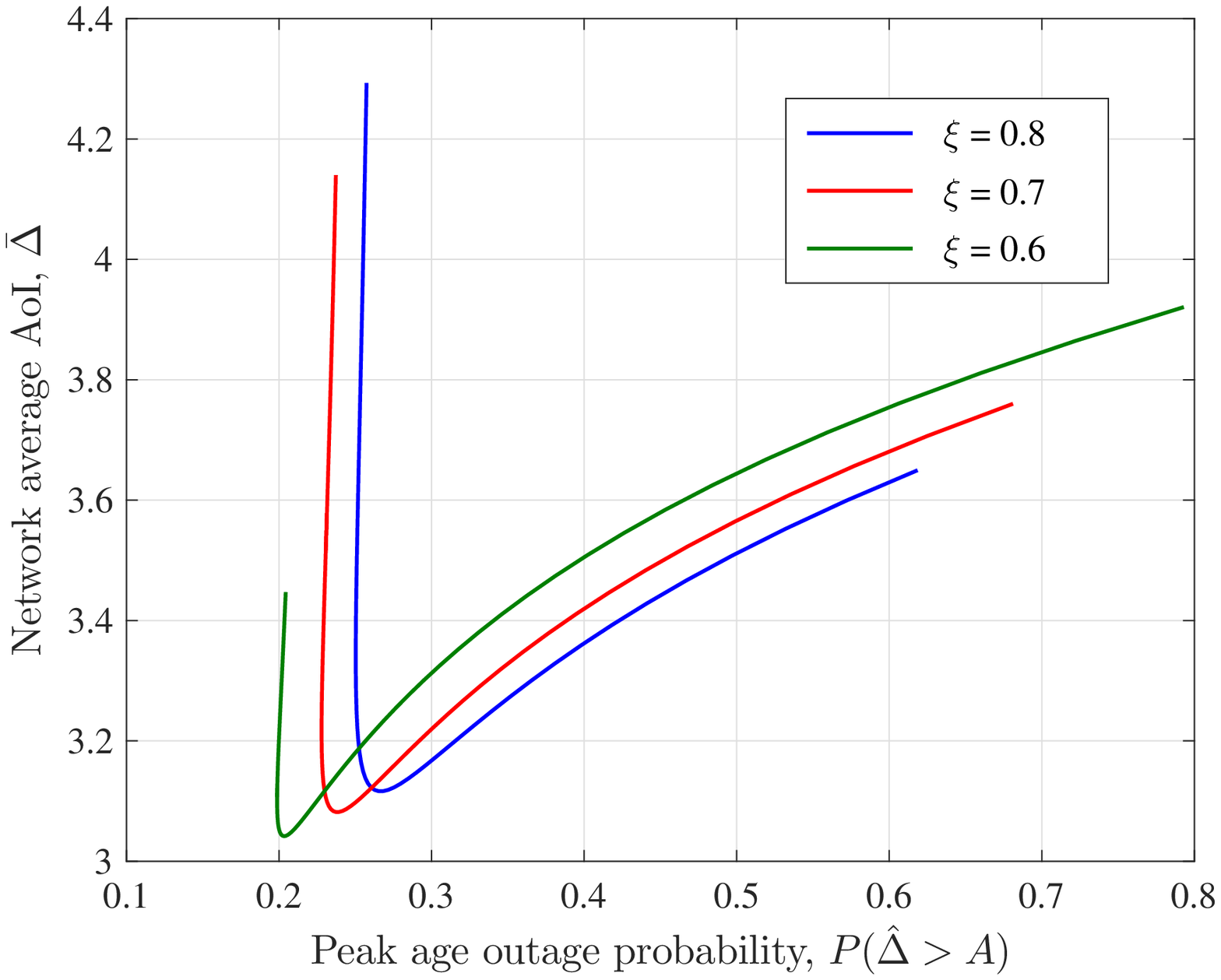}} ~
  \subfigure[\label{fig:1b}]{\includegraphics[width=0.95\columnwidth]{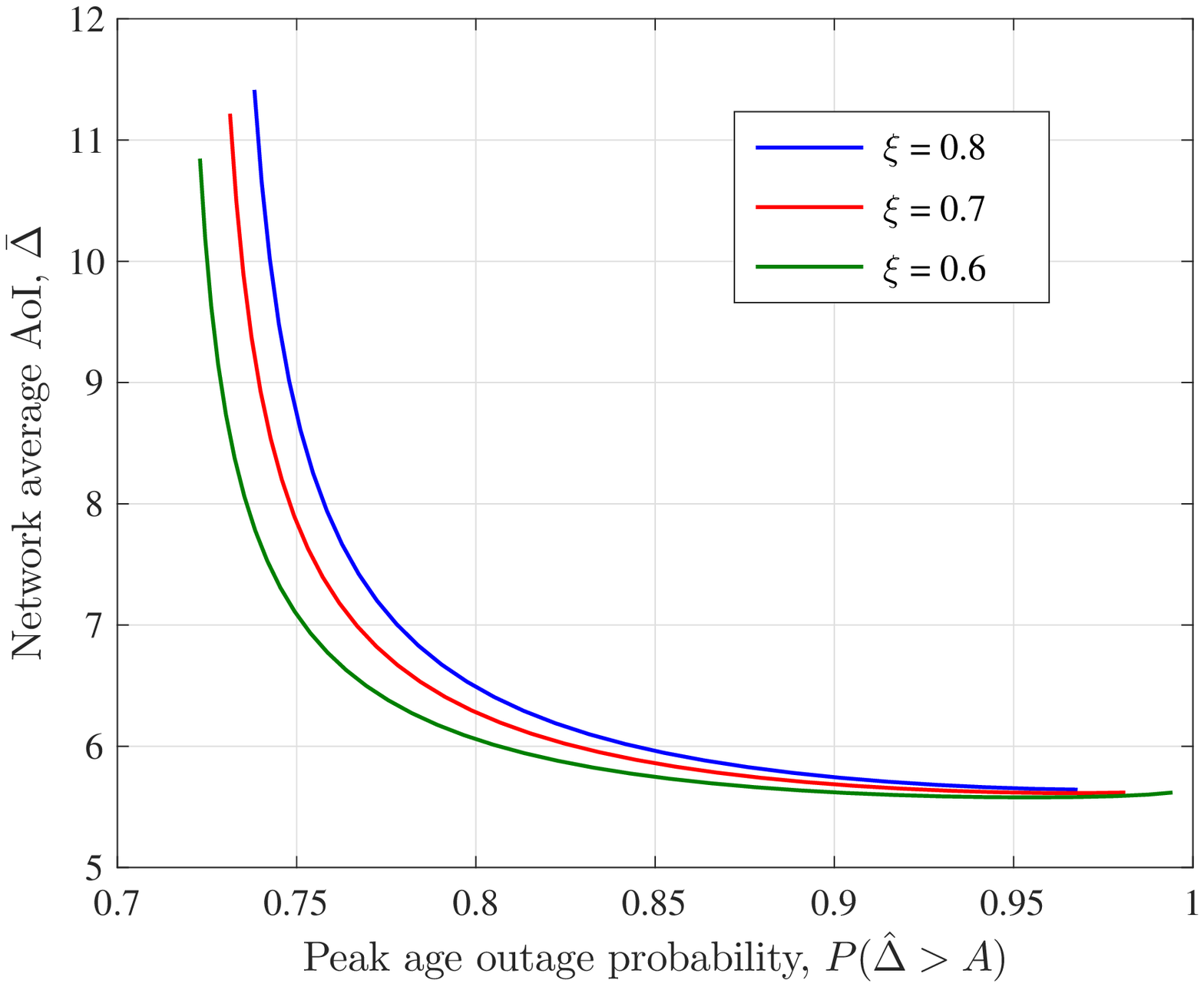}}
  \caption{ Network average AoI versus peak age outage probability: $r=2.5$ m, $p$ increases from 0.4 to 1, and $\xi$ varies as $\xi = 0.6, 0.7, 0.8$. In Fig. (a), the deployment density is $\lambda = 2 \times 10^{-2}$ m$^{-2}$. In Fig. (b), the deployment density is $\lambda = 5 \times 10^{-2}$ m$^{-2}$. }
  \label{fig:AveAoI_vs_Peak}
\end{figure*}

Finally, Fig.~\ref{fig:AveAoI_vs_Peak} compares the network average AoI to the outage probability of peak AoI for a varying value of channel access probability, $p$, under different deployment densities.
From this result, we can see that the average AoI and the outage probability of peak AoI constitute Pareto-like frontiers, which convey two messages: ($a$) if the wireless links are sparsely deployed (i.e., Fig.~\ref{fig:AveAoI_vs_Peak} (a)), there exists an optimal $p$ that minimizes both the average and peak AoI, while ($b$) in densely deployed networks  (i.e., Fig.~\ref{fig:AveAoI_vs_Peak} (b)), one cannot find a channel access probability that simultaneously minimize the average AoI and the outage probability of peak AoI.

\section{Conclusion}
In this work, we established a theoretical framework for the understanding of AoI performance in random access networks. We used a general model that accounts for the channel gain and interference, dynamics of status updating, and spatially queueing interactions.
Our result confirmed that the network topology has a direct and sweeping influence on the AoI.
Specifically, even when the transmitters employ a LCFS-R strategy for packet management, if the topology is densely deployed then there exists an optimal rate of packet arrival that minimizes the average AoI. In addition, slotted ALOHA is instrumental to further reduce the AoI, given the packet arrival rates are high.
However, when the network deployment density is low, the average AoI decreases monotonically with the packet arrival rate, and slotted ALOHA cannot contribute to reducing the AoI in this scenario.
We also found that while similar phenomena also occur in the outage probability of peak AoI, this quantity is more sensitive to the variants of network parameters.
Using the analytical framework, we further developed a channel access policy that configures the probability of channel access at each transmitter based on its observed local information, and hence can be implemented in a decentralized manner and has low complexity.
The proposed scheme can effectively reduce the network average AoI, especially when the network grows in size, as it is able to adaptively adjust according to the spatiotemporal change of the ambient environment.

The analysis developed in this paper manages to straddle queueing theory with stochastic geometry, and allows one to investigate the impacts spatial and temporal factors on the AoI performance. In consequence, this work opens up many exciting directions for future investigation, including but not limited to exploring the impact of different buffer sizes,
retransmission schemes, or channel access approaches on the
AoI of a large-scale wireless network. Investigating up to what
extent power control can improve AoI is also regarded as a
concrete direction for future work.

\begin{appendix}
\subsection{Proof of Lemma~\ref{lma:Cndt_AoI} } \label{apn:Cndt_AoI}
Let us consider a Geo/Geo/1 queueing system under the LCFS with preemption (LCFS-PR) discipline \cite{TriTalMod:19},
where the arrival and departure rates are set as $\xi$ and $p \mu^\Phi_0$, respectively. The AoI in this system evolves as follows:
\begin{align}\label{equ:AoI_evl_LCFS-PR}
\tilde{\Delta}_0(t \!+\! 1) =
\left\{
       \begin{array}{ll}
         \!\!  \tilde{\Delta}_0(t)  +   1, \qquad \qquad  ~~   \text{if transmission fails}, \\
         \!\!  \min\{ t  -  G_0(t), \tilde{\Delta}_0(t) \} + 1, \quad  ~~ \text{otherwise}
       \end{array}
\right.
\end{align}
where $G_0(t)$ is the generation time of the packet delivered over the typical link at time $t$.
From \eqref{equ:AoI_evl_LCFS-PR}, it can be seen that the AoI under the LCFS-PR protocol in a Geo/Geo/1 queue drops only when a more recently generated packet is received at the destination, and that is equivalent to discarding the older undelivered packets at the source. Therefore, this system and the employed system in this paper possess the same AoI evolution statistics. 

We denote by $M$ and $N$ the inter-arrival time and the total sojourn time in the queue, respectively, which are random variables.
As such, under the LCFS-PR discipline, the average AoI is given as \cite{TriTalMod:19}:
\begin{align} \label{equ:Gnl_AoI_L}
\mathbb{E}^0\big[ \bar{\Delta} \vert \Phi \big] = \frac{1}{2} \cdot \frac{ \mathbb{E}\big[ M^2 \big] }{ \mathbb{E}[ M ] } + \frac{\mathbb{E} \big[ \min(N, M) \big] }{ \mathbb{P}\big( N \leq M \big) } - \frac{1}{2}.
\end{align}
On the one hand, as $M \sim Geo(\xi)$ and $N \sim Geo( p \mu^\Phi_0)$, we have the following
\begin{align} \label{equ:Mmnt_M}
&\mathbb{E}[M] = \frac{1}{\xi}, \quad
\mathbb{E}[M^2] = \frac{ 2 - \xi }{ \xi^2 }, \\ \label{equ:Prob_NM}
& \mathbb{P}( N \leq M ) = 1 - \mathbb{E}\big[ ( 1 - p \mu^\Phi_0 )^M \big]
\nonumber\\
&\qquad \qquad ~~ = \frac{\mu^\Phi_0}{ 1 - ( 1 - p \mu^\Phi_0 ) ( 1 - \xi ) }.
\end{align}
On the other hand, since $M$ and $N$ are independent random variables, through simple calculations we have $\min(M,N) \sim Geo( 1 - (1-p\mu^\Phi_0)(1-\xi))$. Thus the following holds
\begin{align} \label{equ:Mmnt_min}
\mathbb{E}\big[ \min( M,N ) \big] = \frac{1}{ 1 - ( 1 - p \mu^\Phi_0 )(1-\xi) }.
\end{align}
The result in \eqref{equ:Cnd_AoI_LCFS} then follows from substituting \eqref{equ:Mmnt_M}, \eqref{equ:Prob_NM}, and \eqref{equ:Mmnt_min} into \eqref{equ:Gnl_AoI_L}.

Next, in a Geo/Geo/1 queue, the conditional peak AoI under LCFS-PR is given as \cite{TriTalMod:19}:
\begin{align} \label{equ:Gnl_PAoI_L}
\mathbb{E}^0\big[ \hat{\Delta} \vert \Phi \big] = \frac{ \mathbb{E} [ M ] }{ \mathbb{P}\big( N \!\leq\! M \big) } + \frac{ \mathbb{E} \big[ N \mathbbm{1}\{ N \!\leq\! M \} \big] }{ \mathbb{P}\big( N \!\leq\! M \big) } - 1.
\end{align}
The numerator of the second term on the R.H.S. above can be calculated as
\begin{align} \label{equ:Epct_NleqM}
 & \mathbb{E} \big[ N \mathbbm{1}\{ N \leq M \} \big] = \mathbb{E}\Big[ \mathbb{E} \big[ N \mathbbm{1}\{ N \leq M \} \big\vert M \big]  \Big]
\nonumber\\
&= \mathbb{E}\Big[ \sum_{m=1}^{M} m ( 1 - p \mu^\Phi_0 )^{m-1} p \mu^\Phi_0  \Big]
\nonumber\\
&= \mathbb{E} \Big[ \, \frac{ 1 - (1 - p \mu^\Phi_0 )^M }{\mu^\Phi_0} - (1-p \mu^\Phi_0) M ( 1 - p \mu^\Phi_0 )^{M-1} \Big]
\nonumber\\
&= \frac{ p \mu^\Phi_0  }{ \big[ 1 - ( 1 - p \mu^\Phi_0 )( 1 - \xi ) \big]^2 }.
\end{align}
The expression in \eqref{equ:Cnd_PAoI_LCFS} then follows by substituting \eqref{equ:Mmnt_M}, \eqref{equ:Prob_NM}, and \eqref{equ:Epct_NleqM} into \eqref{equ:Gnl_PAoI_L}.

\subsection{Proof of Lemma~\ref{lma:Cnd_SucProb} } \label{apn:Cnd_SucProb}
By conditioning on the spatial realization $\Phi$ of the node locations, the transmission success probability can be derived as:
\begin{align}
&\mathbb{P}\big( \mathrm{SINR}_0 > \theta \,|\, \Phi \big)
\nonumber\\
&= \mathbb{P} \Big( \frac{ P_{\mathrm{tx}} H_{00} r^{-\alpha} }{ \sum_{j \neq 0} P_{\mathrm{tx}} H_{j0} \zeta_j \nu_j \Vert X_j \Vert^{ - \alpha } + \sigma^2 } > \theta \,\big|\, \Phi \Big)
\nonumber\\
&= \mathbb{P} \Big( H_{00} > \sum_{ j \neq 0 } \frac{ \nu_j \zeta_j H_{j0} }{ \Vert X_j \Vert^\alpha / \theta r^\alpha } + \frac{ \theta r^\alpha }{ \rho } \big| \Phi \Big)
\nonumber\\
&= e^{ - \frac{ \theta r^\alpha }{ \rho } } \cdot \mathbb{E} \Big[ \prod_{ j \neq 0 } \! \exp\!\Big(\! - \frac{ \nu_j \zeta_j H_{j0} }{ \Vert X_j \Vert^\alpha \! / \theta r^\alpha }  \Big) \big| \Phi \Big]
\nonumber\\
&\stackrel{(a)}{=}  e^{ - \frac{ \theta r^\alpha }{ \rho } } \cdot \prod_{ j \neq 0 } \bigg( 1 - \mathbb{P}\big( \zeta_j \!=\! 1, \nu_j = 1 | \Phi \big)
\nonumber\\
& \qquad \qquad \qquad+ \frac{ \mathbb{P}\big( \zeta_j \!=\! 1, \nu_j = 1 | \Phi \big) }{ 1 + \theta r^\alpha \! / \Vert X_j \Vert^\alpha } \bigg)
\nonumber\\
&\stackrel{(b)}{=}  e^{ - \frac{ \theta r^\alpha }{ \rho } } \cdot \prod_{ j \neq 0 } \bigg( 1 - \mathbb{P}\big( \zeta_j \!=\! 1 | \Phi \big) \times \mathbb{P}\big( \nu_j = 1 | \Phi \big)
\nonumber\\
& \qquad \qquad \qquad+ \frac{ \mathbb{P}\big( \zeta_j \!=\! 1 | \Phi \big) \times \mathbb{P}\big( \nu_j = 1 | \Phi \big) }{ 1 + \theta r^\alpha \! / \Vert X_j \Vert^\alpha } \bigg)
\end{align}
where ($a$) follows from Assumption~1, under which the active state of each node can be regarded as independent, and notice that $H_{j0} \sim \exp(1)$. The result can then be obtained via further simplifying the product factors.

\subsection{Proof of Lemma~\ref{lma:Cndt_ActProb} } \label{apn:Cndt_ActProb}
The evolution of the buffer state at a generic node $j$ can be modeled as a two-state Markov chain (empty/non-empty) with transition matrix given as follows:
\begin{align}
\mathbf{P} \!=\!
  \begin{bmatrix}
    ~1 - \xi &   \xi
    \nonumber\\
    ~ p \mu^\Phi_j (1-\xi) &  1 - p\mu^\Phi_j + p\mu^\Phi_j \xi
   \end{bmatrix}.
\end{align}
Let $\mathbf{v} = (v_0, v_1)$ denote the steady-state probability vector of the number of this Markov chain. Then, we have
\begin{align}
\mathbf{v}^{\mathrm{T}} = \mathbf{v}^{\mathrm{T}} \mathbf{P}, \\
v_0 + v_1 = 1.
\end{align}
Solving the above system of equations yields the following:
\begin{align}
v_0 = \frac{ p \mu^\Phi_j (1-\xi) }{ \xi + p \mu^\Phi_j (1-\xi) }, \\ \label{equ:buff_1}
v_1 = \frac{ \xi }{ \xi + p \mu^\Phi_j (1-\xi) }.
\end{align}
As such, the active state probability $a^\Phi_j$ can be obtained from \eqref{equ:buff_1} (the probability of having a non-empty buffer).

\subsection{Proof of Theorem~\ref{thm:Meta_SINR} } \label{apn:Meta_SINR}
For ease of exposition, let us denote $Y^\Phi_0 = \ln \mathbb{P}(\mathrm{SINR}_0 > \theta | \Phi)$. By leveraging Lemma~\ref{lma:Cnd_SucProb} and Lemma~\ref{lma:Cndt_ActProb}, we can calculate the moment generating function of $Y^\Phi_0$ as follows
\begin{align} \label{equ:MY_moment}
&\mathcal{M}_{ Y^\Phi_0 } (s) = \mathbb{E}\big[ ( \mu^\Phi_0 )^s \big]
\nonumber\\
& = e^{ - \frac{s \theta r^\alpha}{ \rho } } \mathbb{E}\Big[ \prod_{ j \neq 0 } \! \big( 1 - \frac{ a^\Phi_j }{ 1 \!+\! \Vert X_j \Vert^\alpha / \theta r^\alpha } \big)^s \Big]
\nonumber\\
& = e^{ - \frac{s \theta r^\alpha}{ \rho } } \mathbb{E}\Big[ \prod_{ j \neq 0 } \! \big( 1 - \frac{ 1 }{ 1 \!+\! \Vert X_j \Vert^\alpha / \theta r^\alpha } \cdot \frac{ p \xi }{ \xi \!+\! ( 1 \!-\! \xi ) p \mu^\Phi_j } \big)^s \Big]
\nonumber\\
& \stackrel{(a)}{=} e^{ - \frac{s \theta r^\alpha}{ \rho } } e^{- \lambda \int_{ \mathbb{R}^2 } \mathbb{E} \Big[ 1 - \big( 1 - \frac{1}{ 1 + \Vert x \Vert^\alpha / \theta r^\alpha } \cdot \frac{ p \xi }{ \xi + ( 1 - \xi ) p \mu_x } \big)^s \Big] dx }
\nonumber\\
& \stackrel{(b)}{=} \exp \! \bigg(\! - \frac{ s \theta r^\alpha }{ \rho } - \lambda \! \int_{ \mathbf{x} \in \mathbb{R}^2 } \sum_{ k = 1 }^{s} \! \binom{ s }{ k }   \frac{ (-1)^{k+1} d \mathbf{x} }{ ( 1 \!+\! \Vert \mathbf{x} \Vert^\alpha \!/ \theta r^\alpha )^k }
\nonumber\\
&\qquad \qquad \qquad \qquad \qquad \qquad \times \underbrace{\mathbb{E} \Big[ \big( \frac{ p \xi }{ \xi + ( 1 - \xi ) p \mu_{\mathbf{x}} } \big)^k \Big] }_{Q_1} \bigg),
\end{align}
where ($a$) follows by using the probability generating functional (PGFL) of PPP and ($b$) expands the expression via the binomial theorem.
Note that under Assumption~1, the conditional transmission success probability $\mu_{\mathbf{x}}$ can be considered as i.i.d. across the transmitters.
A complete expression of \eqref{equ:MY_moment} requires us to compute the term $Q_1$, which however needs the CDF, $F(\cdot)$, of $\mu^\Phi_0$.
At this stage, let us assume the function $F(\cdot)$ is available. We can then expand the expectation term $Q_1$ and further reduce \eqref{equ:MY_moment} as we do below:
\begin{align} \label{equ:MY_reduce}
\mathcal{M}_{ Y^\Phi_0 }(s) &= \exp\!\bigg(\!\! - \frac{ s \theta r^\alpha }{ \rho } - \!\!\!  \int\limits_{ \mathbf{x} \in \mathbb{R}^2 } \sum_{ k = 1 }^{s} \! \binom{ s }{ k }   \frac{  (-1)^{k+1} \lambda d \mathbf{x} }{ ( 1 \!+\! \Vert \mathbf{x} \Vert^\alpha \! / \theta r^\alpha )^k }
\nonumber\\
&\qquad\qquad \qquad \qquad \times \int_{0}^{1}  \Big( \frac{ p \xi }{ \xi + ( 1 - \xi ) p t } \Big)^k F(dt) \bigg)
\nonumber\\
& = \exp\!\bigg(\!\! - \frac{ s \theta r^\alpha }{ \rho } - \lambda \pi r^2 \theta^\delta \sum_{ k = 1 }^{s} \! \binom{ s }{ k } \frac{ (-1)^{k+1} \! \int_{0}^{\infty} \! dv }{ ( 1 \!+\! v^{ \frac{\alpha}{2} } )^k }
\nonumber\\
&\qquad \qquad \qquad \qquad\qquad~\, \times \int_{0}^{1} \frac{ ( p \xi )^k F(dt) }{ \big[ \xi + ( 1 - \xi ) p t \big]^k }  \bigg).
\end{align}

Finally, by using the Gil-Pelaze theorem \cite{Gil}, we can derive the CDF of $\mu^\Phi_0$ as:
\begin{align}
F(u) &= \mathbb{P}( \mu^\Phi_0 < u ) = \mathbb{P}( Y^\Phi_0 < \ln u )
\nonumber\\
&= \frac{1}{2} - \frac{1}{\pi} \int_{0}^{\infty} \mathrm{Im} \big\{ u^{ - j \omega } \mathcal{M}_{Y^\Phi_0} ( j \omega ) \big\} \frac{d \omega}{ \omega }.
\end{align}
The statement readily follows by substituting \eqref{equ:MY_reduce} into the above equation.

\subsection{Proof of Corollary~\ref{cor:beta_approx} } \label{apn:beta_approx}
According to \cite{YanQue:19}, the fixed-point equation \eqref{equ:Meta_Grl} can be solved by recursively evaluating the following
\begin{align}
&F_{n+1}(u) = \frac{1}{2} -\! \int_{0}^{\infty} \!\!\! \mathrm{Im}\bigg\{ u^{-j\omega} \exp\!\Big(\! - \frac{ j \omega \theta r^\alpha }{ \rho } -  \lambda \pi r^2 \theta^\delta
\nonumber\\
&\times \sum_{k=1}^{\infty}  \binom{j \omega}{ k } \!\int_0^\infty \frac{ (-1)^{k+1} dv }{ (1+v^{ \frac{ \alpha }{ 2 } } )^k }  \int_{0}^1 \!\! \frac{ (p \xi)^k F_n(dt) }{ [\, \xi + (1-\xi) p t \,]^k } \Big) \bigg\} \frac{ d \omega }{ \pi \omega }.
\end{align}
Since the function $F_{n}(u)$ in each iteration step is supported on $[0,1]$, we are motivated to approximate the distribution via a Beta distribution.
First, by assigning $s$ as integers as per \eqref{equ:MY_moment}, we can derive the moments of $\mu^\Phi_0$ in \eqref{equ:Momnt_Beta}. Next, by respectively matching the mean and variance of $\mu^\Phi_0$ to those of a Beta distribution $B(a_n, b_n)$, it yields
\begin{align}
& \frac{ a_n }{ a_n + b_n } = c_n^{(1)}, \\
& \frac{ a_n b_n }{ ( a_n + b_n )^2 ( a_n + b_n + 1 ) } = c_n^{(2)} - \big[ c_n^{(1)} \big]^2
\end{align}
and the result follows from solving the above system equations.

\subsection{Proof of Theorem~\ref{thm:DmSym_PeakAoI}} \label{apx:OptCtrl_PeakAoI}
Following Assumption~1, the point process $\Phi$ can be regarded as stationary under the employed network model. We can thus substitute \eqref{equ:CndTXSucProb_Domint} into the first term of \eqref{prbm:Dom_PeakAoI} and use the mass transportation theorem \cite{BacBla:09} to obtain the following:
\begin{align}\label{equ:MasTrn}
&\mathbb{E}^0_\Phi \!\! \left[ \frac{ \xi }{  \gamma^\Phi_0 \mu^{\Phi | W}_0 } \right] \!=\!  \mathbb{E}^0_\Phi \! \left[ \frac{ \exp\!\left({\frac{ \theta r^\alpha }{\rho}} \!+\!  \int_{ \mathbf{x} \in \mathbb{R}^2 \setminus W }  \frac{ \lambda \eta_{_W} \mathbb{E}\big[ a^\Phi_{\mathbf{x}} \big]  d\mathbf{x} }{ 1 + \Vert \mathbf{x} \Vert^\alpha \!/ \theta r^\alpha } \right) }{ \eta_{_W} \! \prod_{j \neq 0} \! \left( 1 \! - \! \frac{ \eta_{_W} a^\Phi_j }{ 1 + \mathcal{D}_{0j} } \right)  } \right]\!.
\end{align}
It shall be noted that while the buffer states of the transmitters are in fact coupled in both space and time, we leverage Assumption~1 to decouple them so as to arrive at a tractable expression as per \eqref{equ:MasTrn}. Consequently, optimizing \eqref{prbm:minAoI} is now equivalent to minimize the right hand side of above equation, as a function of $\eta_{_W}$, under the constraint in \eqref{equ:Dom_Rst_SpSt}. In general, such a functional optimization should be  solved via the calculus of variants. Since the operator $\eta_{\mathrm{S}}$ is well-defined in the stationary point process $\Phi$, we can treat it as a variable \cite{BacBlaSin:14} and assign the derivative of \eqref{prbm:Dom_PeakAoI} with respect to $\eta_{_{W}}$ to be zero, which yields the following:
\begin{align} \label{equ:FixPnt_PRF}
\frac{1}{\eta_{_{W}}} -\!\!\!\!\!\! \sum_{ \substack{ j \neq 0, y_j \in W } }  \frac{1}{ 1 \!+\! \mathcal{D}_{0j}   \!-\! \eta_{_{W}} a^\Phi_j } -\!\! \int_{\mathbb{R}^2 \setminus W }\! \frac{ \lambda \mathbb{E}[a^\Phi_0]  dz }{ 1 \!+\! z^\alpha \! / \theta r^\alpha } = 0.
\end{align}
If we write the left hand side (L.H.S.) of the above equation as a function $h(\eta_{_{W}})$ of $\eta_{_{W}}$, i.e.,
\begin{align}
h(\eta_{_W}) = \frac{1}{\eta_{_{W}}} -\!\!\!\!\! \sum_{ \substack{ j \neq 0, y_j \in W } } \frac{1}{ 1  \!+\! \mathcal{D}_{0j} \!-\! \eta_{_{W}} a^\Phi_j } -\!\! \int_{\mathbb{R}^2 \setminus W }\! \frac{ \lambda  \mathbb{E}[a^\Phi_0]  dz }{ 1 \!+\! \Vert z \Vert^\alpha \! / \theta r^\alpha },
\end{align}
it is easy to verify that ($a$) $h(\eta_{_W})$ monotonically decreases in $\eta_{_W}$ over [0, 1], and ($b$) $\lim_{ \eta_{_W} \rightarrow 0 } h(\eta_{_W}) = +\infty$. As such, if $h(1) < 0$, i.e., the condition \eqref{equ:CndOptl} holds, then according to the Intermediate Value Theorem, the equation in \eqref{equ:FixPnt_PRF}, or equivalently, (38), has a unique solution that lies within the interval (0, 1).
Otherwise, if \eqref{equ:CndOptl} does not hold, we have the derivative of \eqref{equ:MasTrn} being negative which indicates that \eqref{prbm:Dom_PeakAoI} monotonically decreases as a function of $\eta_{_{W}}$. Hence, the minimum is achieved at $\eta_{_{W}} = 1$. Note that in both cases, the constraint (35b) is satisfied. 

\end{appendix}

\bibliographystyle{IEEEtran}
\bibliography{bib/StringDefinitions,bib/IEEEabrv,bib/howard_AoI_RAN}

\end{document}